\documentclass{ws-ijmpd}
\usepackage{fix-cm}
\usepackage{amssymb,graphicx,amsmath,graphics,xfrac,url,booktabs,adjustbox}
\let\captionbox\relax
\usepackage{caption}
\usepackage{subfigure}
\usepackage{subcaption}
\usepackage{multicol}
\usepackage{multirow}

\newcommand\sr{\mathrm{sr}}
\newcommand\ini{\mathrm{ini}}
\newcommand\en{\mathrm{end}}

\newcommand\sca{\mathrm{S}}
\newcommand\ten{\mathrm{T}}

\newcommand\ret{\mathrm{ret}}
\newcommand\D{\mathrm{d}}
\newcommand\ele{\mathrm{l}}
\newcommand\ere{\mathrm{r}}
\newcommand\e{\mathrm{e}}
\newcommand\num{\mathrm{num}}
\newcommand\phai{\mathrm{pi}}
\newcommand\ua{\mathrm{ua}}
\newcommand\im{\mathrm{i}}

\newcommand\Mpc{\mathrm{Mpc}}

\newcommand\fit{\mathrm{fit}}

\DeclareMathOperator{\arccosh}{arccosh}

\begin{document}
	
\title{\boldmath Quintessential inflation studied through  Semiclassical Methods}

\author{Jordan Zambrano$^{1,3}$, Miguel Agama$^2$,  Marcos Garz\'on$^1$, Werner Br\"amer--Escamilla$^1$, Clara Rojas$^1$ \thanks{crojas@yachaytech.edu.ec}, and  Te\'ofilo Vargas$^2$}

\address{$^1$ Yachay Tech University, School of Physical Sciences and Nanotechnology, Hda. San Jos\'e S/N y Proyecto Yachay, 100119, Urcuqu\'i, Ecuador}

\address{$^2$ Grupo de F\'isica Te\'orica and Grupo de Astronom\'ia SPACE, Universidad Nacional Mayor de San Marcos,
Avenida Venezuela S/N Cercado de Lima, 15081, Lima, Per\'u}

\address{$^3$ Universidad de Chile, Departamento de F\'isica, FCFM, Blanco Encalada 2008, Santiago, Chile}

\maketitle
	
\begin{history}
\received{\today}
\end{history}

\begin{abstract}

In this work, we solved the scalar and tensor perturbation equations numerically and using the improved uniform approximation method together with the third--order phase--integral method, for the $\alpha$--attractor inflationary model. This inflationary model has become very important because it allows us to describe the initial accelerated expansion of the universe in the inflationary epoch, and the current accelerated expansion with the same potential that depends on one scalar field $\varphi$. Once the equations for the scalar and tensor power spectra are found, we calculate the observables: the scalar--to--tensor ratio $r$, and the scalar spectral index $n_\sca$, concluding that semiclassical methods give excellent results compared to numerical integration. We also compare both observables in the $\alpha$--attractor and the Starobinsky inflationary model.

\end{abstract}

\maketitle
\section{Introduction}

Inflationary epoch has become a central part of modern cosmology since it provides the solution to the horizon, the flatness, the monopole problems, and also has the property of producing scalar cosmological perturbations which seeds the structure formation and the anisotropies of the Cosmic Microwave Background Radiation (CMB), whereas the tensor cosmological perturbations produces primordial gravitational waves (for reviews on inflation see e.g., \cite{linde1984inflationaryONE,baumann2009tasiTWO,martin2017theoryFOUR,martin:2014}). 

Scalar field, known as the inflaton, plays a fundamental role in cosmic inflation because it acts as the driving force behind the universe exponential expansion, and the inflaton quantum fluctuations during this early epoch, stretched to cosmological scales, are the seeds for all cosmic structures, evolving into the large--scale distribution of matter we observe today \cite{martin:2014}.

On the other hand, the cosmological observations indicate that the universe is currently experiencing an accelerated expansion \cite{riess1998observational,Perlmutter1999,Li2013,Abbott2023}. This fact can be attributed to the existence of some form of energy (dubbed Dark Energy (DE)) with negative pressure effectively acting as a cosmic repulsive force that pushes galaxies apart at an increasing rate.
Thus,  DE is now a key component in the $\Lambda$-CDM model ($\Lambda CDM$), the standard cosmological theory that describes a universe dominated by dark energy ($\Lambda$), cold dark matter (CDM), and ordinary matter \cite{Abbott2023}.
Consequently, according to the modern cosmological model there are two accelerated phases of our universe, and both phases are generally treated independently. Furthermore,  in $\Lambda$CDM model, in order for the cosmological constant energy density to be dominant today, it must have contributed a negligible amount to the total energy density during recombination, reheating and nucleosynthesis. This fact led to the idea of a time variable cosmological constant  which can potentially account for the dark energy in a dynamical way \cite{Peebles1988,Coble1997,Lopez1996,Linde1995}.

One generalization of a dynamical cosmological constant is the time--dependent and spatially inhomogeneous component with a negative pressure scalar field, whose equation--of--state is different from baryons, neutrinos, dark matter, and photons \cite{Peebles1988,Wetterich1988,Caldwell1998,Tsujikawa2013}: Quintessence.
However, it was soon realized that entangling inflation and quintessence is rather natural because, according to both theories, the Universe undergoes a phase of accelerated expansion when dominated by the energy density of a scalar field, which slowly rolls down its almost flat potential. Such a scenario is known as quintessential inflation \cite{PeeblesVilenkin1999,Sahni2001,DimopoulosValle2002,Hossain2015,deHaro2021,BettoniRubio2022}. The quintessence during inflation is at very high--energy, rapid expansion of the universe, while present--day quintessence is a slowly evolving scalar field responsible for the later lower--energy, accelerating expansion we observe now. Furthermore,
the quintessential field should not interfere with the thermal history of the universe, and it would reappear only recently, giving rise to late--time cosmic acceleration. This requires that the scalar field potential have a steep region in the potential just after the end of inflation. 

In recent years, generalized $\alpha$--attractor inflationary models have attracted considerable attention because they have not only a well--understood origin in conformal and super--conformal field theories but also they have a close relationship with supergravity theories \cite{Kallosh_Linde:2013b,Kallosh_Linde_Roest:2013,galante,carrasco2015,carrasco2016}. This means that there are also connections between these models and string theory, with some models being constructed using K\"ahler and super potentials that appear natural from a string theory perspective \cite{ferrara,kallosh8}. 
Furthermore, their general predictions lead to a general prediction for inflationary observables and lie close to the center of current observational bounds on the primordial power spectra, namely the scalar spectral index $n_S(k)$ and the tensor--to--scalar ratio $r(k)$ which are consistent with observations from the Planck satellite \cite{akrami:2020,rodrigues2021observational,bhatta}.

The most recent exciting tension in cosmology is the Hubble tension, arising from a growing discrepancy between the $\Lambda CDM$--based predictions of the Hubble constant from cosmic microwave background measurements and direct measurements in the local universe.
One of the most promising candidates to resolve this tension is the inclusion of an Early Dark Energy (EDE) component that can inject energy near the time of recombination, diluting quickly afterward to allow the EDE to smoothly transition into the observed late--time dark energy, thus alleviating the Hubble tension \cite{PettorinoAmendolaWetterich2013,KarwalKamionkowski2016,PoulinSmithKarwalKamionkowski2019,sabla}.  The $\alpha$--attractor inflationary models provide a unified theoretical framework that can naturally accommodate both EDE and late--time dark energy within the same scalar field potential \cite{BragliaEmondFinelliGumrukcuogluKoyama2020,BrissendenDimopoulosSanchezLopez2024,sarkar}.

In this work, we develop an important type of model in inflationary cosmology, the $\alpha$-attractor inflationary model. Its importance comes from the relation this model provides between the theoretical justification of the inflationary epoch and late--time cosmic acceleration. Describing inflation and dark energy dynamics with the same approach \cite{salo:2021,akrami2021quintessential}.
This model is also part of quintessential inflation models, which offer an approach that lets relate the physics of inflation, that in $\Lambda$CDM model is represented by a scalar field minimally coupled to gravity driven by a potential, and late--time acceleration, manifested through DE \cite{linder2015dark}, usually included in the form of a positive cosmological constant in Einstein's field equations in the framework of the standard cosmological model.
This model offers an alternative approach to solving some tensions regarding cosmological parameters as mentioned before, proposing an asymptotically de Sitter Universe, in contrast to what is expected from extensions of fundamental physics like the quantum gravity approach that considers an asymptotically anti--de Sitter Universe \cite{giare2024testing}.
In that sense, $\alpha$--attractor inflationary models offer an optimistic path to obtain quintessential inflation, considering that it follows some properties that make it a favorable option like: its stability and robustness against quantum corrections, its attractor behavior that increases its independence from initial conditions, and also the implication it has referring to the dynamics of DE described in a way distinct from a cosmological constant.

Quintessential inflation offers a new representation for dark energy in the form of a dynamical and time--dependent equation of state parameter $\omega_{0}$. And through a consistency relation, this term can be related to cosmological parameters obtained from inflation, tensor--to--scalar ratio $r$, and number of e--foldings $N$ \cite{zhumabek2023connecting},

\begin{equation}
\omega_{0} = -1 + \dfrac{4}{3N^2\, r},
\label{alpha_1}
\end{equation}
where $\omega_{0}$ is the current value of the parameter, and also this value is related to the function $\omega(a)$, which varies with respect to the scale factor \cite{de2008calibrating},

\begin{equation}
\omega(a) = \omega_{0} + \omega_{a}(1-a),
\label{alpha_2}
\end{equation}
being $\omega_{a}$ treated as a fit parameter from observational measurements. This is in contrast to the constant value $\omega = -1$ that is obtained from the cosmological constant approach \cite{yadav2011dark}.

For early times, we calculate the slow--roll parameters that are given by \cite{liddle2000cosmological},

\begin{eqnarray}
\label{epsilon}
\epsilon &\simeq& \dfrac{1}{2} \left(\dfrac{V'}{V}\right)^2,\\
\label{eta}
\eta &\simeq& \dfrac{V''}{V},
\end{eqnarray}
where prime means derivative with respect to the scalar field $\varphi$.

Using the slow--roll parameters we can calculate the scalar spectral index $n_\sca(k)$ and the scalar--to--tensor ratio $r(k)$,

\begin{eqnarray}
\label{nS_initial}
n_\sca &\simeq&1-6\epsilon+2\eta, \\
\label{r_initial}
r &\simeq & 16 \epsilon.
\end{eqnarray}

In order to calculate the value of the scalar field at the end of inflation $\varphi_\en$ we made $\epsilon=1$. To calculate the value of the scalar field at the beginning of inflation $\varphi_\ini$, we need to calculate the number of e--folding using \cite{liddle2000cosmological},

\begin{equation}
\label{Ne}
N \simeq \int_{\varphi_\en}^\varphi \dfrac{V}{V'} \D \varphi.
\end{equation}

This paper is organized in the following way: in Section \ref{Model}, we introduce the $\alpha$--attractors inflationary model. Section \ref{Background} presents the numerical solution of the background equations using the slow--roll approximation. In this section, we also fit the numerical solution for the scale factor $a(t)$ and the scalar field $\varphi(t)$. Section \ref{Perturbations} describes the scalar and tensor perturbation equations, which are later solved through four different solution methods in Section \ref{Solution_Perturbations}. These methods are applied to our model in Section \ref{Methods}, followed by the presentation of our results in Section \ref{Results}. Finally, our  conclusions are presented in Section \ref{Conclusions}.

\bigskip
\section{The $\alpha$--attractor Inflationary Model}
\label{Model}

\bigskip
A brief overview of the theory of the $\alpha$--attractors inflationary models is provided here. We consider the following non--standard Lagrangian density, motivated by supergravity and corresponding to a nontrivial K\"ahler manifold \cite{galante}, combined with an exponential potential in four--dimensional spacetime \cite{salo:2021}

\begin{equation}
\label{lagrangian}
\mathcal{L} = \dfrac{\frac{1}{2}g^{\mu\nu}\partial_{\mu}\phi\partial_{\nu}\phi}{(1-\sfrac{\phi^2}{6\alpha})^2}-\lambda e^{-k\phi},
\end{equation}
where $\phi$ is the inflaton field, $\alpha>0$ the free parameter of the theory related to free curvature of the K\"ahler manifold, $\alpha$ is the parameter determining  the steepness of the potential  and, $\lambda$ is a constant with dimension of energy density.

The primary characteristics of these models are evident due to the presence of a non-canonical kinetic term for the inflaton, which features poles at $\phi = \pm \sqrt{6\alpha}$. These poles prevent any displacement in the field space from crossing this limit, thus imposing a bound on its value.

One can get the canonical form of the kinetic term defining the non--canonical field $\phi$ in terms of the canonical scalar field $\varphi$ by solving the equation,

\begin{equation}
\partial\varphi=\dfrac{\partial\phi}{(1-\sfrac{\phi^2}{6\alpha)}},
\end{equation}
with  the solution $\phi=\sqrt{6\,\alpha}\,\tanh{\left(\frac{\varphi}{\sqrt{6\alpha}}\right)}.$

\begin{figure}[th!]
\centering
\includegraphics[scale=0.5]{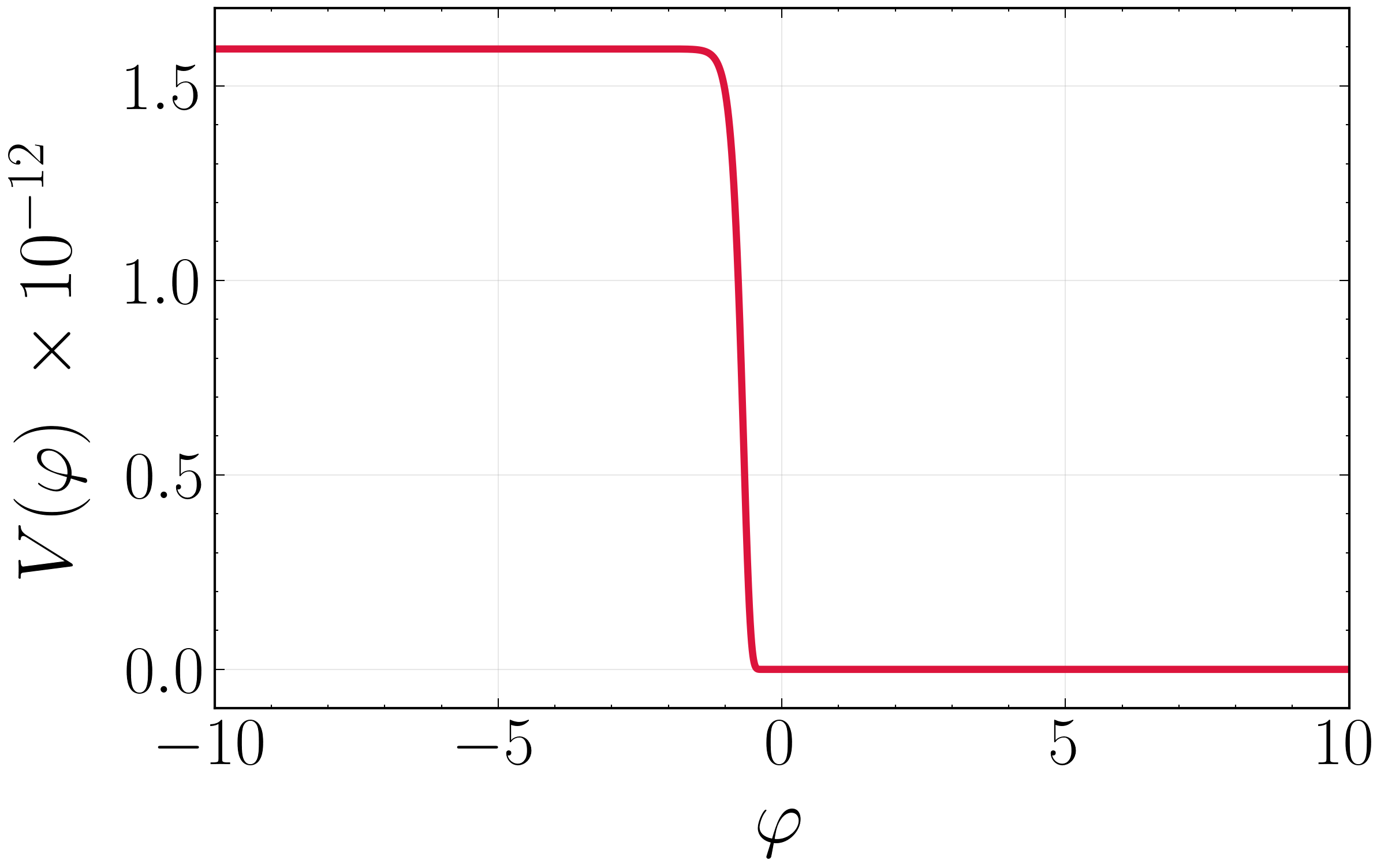}
\caption{$\alpha$--attractor inflationary potential for $n=124$,  
$\alpha=10^{-2}$, and $\lambda=2.2399 \times 10^{-66}$.}
\label{V_alpha_python}
\end{figure}

By this field redefinition, the Lagrangian density of the canonical form is,

\begin{equation}
\mathcal{L}=-\dfrac{1}{2}g^{\mu\nu}\partial_{\mu}\varphi\,\partial_{\nu}\varphi-V(\varphi), 
\end{equation}
and the scalar potential is
\begin{equation}
\label{model}
V(\varphi)=\lambda \, e^{-n\tanh\left(\dfrac{\varphi}{\sqrt{6\alpha}}\right)},
\end{equation}
where $n$, and $\alpha$ are dimensionless parameters, and $n$ depends on $\alpha$ in the following way $n=\kappa\,\sqrt{6\alpha}$. \textbf{The parameters $n$ and $\alpha$ are set
by the authors in } Sal\'o in $n = 124$ and $\alpha=10^{-2}$ \cite{salo:2021}. For these parameters, the form of the $\alpha$--attractor potential Eq. \eqref{model} becomes, see Figure \ref{V_alpha_python},

\bigskip

Here, it is important to remark that changing to a canonical field $\varphi$, transposes the pole to infinity while the scalar potential of the canonical field is stretched near the pole, which generates a plateau in the potential \cite{kallosh3}. Specifically, the inflationary regime is featured by a plateau $\phi \rightarrow-\sqrt{6\alpha}$ or by $\varphi\rightarrow-\infty$, and the quintessence is featured by the often plateau $\phi\rightarrow\sqrt{6\alpha}$ or by $\varphi\rightarrow+\infty$, see Figure \ref{V_alpha_python}.
This potential comes from a generalization of some models that were found to follow some common characteristics about observable quantities' behavior \cite{Kallosh_Linde:2013}. Among these models were Starobinsky and modified versions of it, Higgs, and other inflationary potentials whose predictions regarding scalar spectra index $n_\sca(k)$, and tensor--to--scalar ratio $r(k)$ follow a similar pattern \cite{canas:2021}: which is said to be the universal attractor regime \cite{Kallosh_Linde_Roest:2013}, that this family of potentials tends to reach for small values of $\alpha$ under the context of conformal and superconformal symmetry \cite{Kallosh_Linde:2013b}. Following this, it is possible to recover the case of Starobinsky inflation when the value of $\alpha$ is fixed to $1$, and for large values of the field \cite{salo:2021,canas:2021,Miranda_Fabris_Piattella:2017}.


For this model, the slow--roll parameters are calculated from Eqs. \eqref{epsilon} and \eqref{eta}, and are given by,

\begin{eqnarray}
\label{epsilon_alpha}
\epsilon &\simeq& \dfrac{n^2}{12 \, \alpha} \dfrac{1}{\cosh^4\left(\dfrac{\varphi}{\sqrt{6\alpha}}\right)},\\
\label{eta_alpha}
\eta &\simeq& \dfrac{n}{3\,\alpha}\left[ \dfrac{
n}{2\,\cosh^4\left(\dfrac{\varphi}{\sqrt{6\alpha}}\right)}+\dfrac{
\tanh\left(\dfrac{\varphi}{\sqrt{6\alpha}}\right)}{\cosh^2\left(\dfrac{\varphi}{\sqrt{6\alpha}}\right)}\right]. 
\end{eqnarray} 


Using Eqs. \eqref{epsilon_alpha} and \eqref{eta_alpha}, it is possible to approximate them to the results given by Starobinsky model, with $\alpha = 1$,

\begin{equation}
\epsilon \simeq  \dfrac{4 \, n^2}{3} \left( \frac{1}{e^{\sfrac{\varphi}{\sqrt{6}}} + e^{-\sfrac{\varphi}{\sqrt{6}}}}\right)^4.
\end{equation}

In the limit for $|\varphi|\gg1$,

\begin{equation}
\epsilon \simeq  \frac{4 \,n^2}{3} e^{-2 \sqrt{\sfrac{2}{3}}\, \varphi},
\end{equation}
that follows the same expression of the $\epsilon$ slow-roll parameter in the Starobinsky inflationary model in \cite{truman:2020} at the same limit. For the second slow--roll parameter $\eta$, we can obtain the same match following a similar behavior,

\begin{eqnarray}
\nonumber
\eta &\simeq& \frac{n}{3} \left[ \frac{ 2^{4}n}{2\left( e^{\varphi/\sqrt{6}} + e^{-\varphi/\sqrt{6}}\right)^{4}} + \frac{4\left( e^{\varphi/\sqrt{6}} - e^{-\varphi/\sqrt{6}}\right)}{\left( e^{\varphi/\sqrt{6}} + e^{-\varphi/\sqrt{6}}\right) \left( e^{\varphi/\sqrt{6}} + e^{-\varphi/\sqrt{6}}\right)^{2}} \right],\\
&\simeq& \frac{4 n}{3} e^{-2\varphi/\sqrt{6}}.
%
\end{eqnarray}
This shows a direct relation between these models. Next we will detail how close the behavior is of both models, not only under slow roll regime. Meanwhile, we detail some other features of this approach. Moreover, it can be understood as a family of models. Where this $\alpha$--attractor inflationary model, in  the leading order of $N$ for $\alpha\ll1$ and $N\gg1$, leads to the following prediction for the scalar spectral index and the tensor--to--scalar ratio into the slow--roll approximation \cite{Kallosh_Linde:2013b,salo:2021},  see Eqs. \eqref{nS_initial} and \eqref{r_initial}.

\begin{eqnarray}   
n_\sca &\simeq&  1-\frac{2}{N},\\
r     &\simeq& \frac{2\alpha}{N^2},
\end{eqnarray}
This is said to be the universal attractor regime \cite{Kallosh_Linde_Roest:2013}, that this family of potentials tends to reach  under the context of conformal and superconformal symmetry \cite{Kallosh_Linde:2013b}. Following this, it is possible to recover the case of Starobinsky inflation when the value of $\alpha$ is fixed to $1$ \cite{salo:2021,canas:2021, Miranda_Fabris_Piattella:2017}. Note that this results are compatible not only with WMAP \cite{hinshaw2013nineDOCE}, Planck \cite{ade2016planckTRECE}, and \cite{akrami2021quintessential,iacconi2023novelQUINCE,iacconi2025testingDISIX} cosmic microwave background (CMB) observations, but also with recent DESI collaboration results \cite{alestas2025desiSEVEN}.

Now, let us determine the value of the scalar field 
$\varphi$ at the end of inflation for the $\alpha$--attractor inflationary model. Setting 
$\epsilon=1$ in,  Eq. \eqref{epsilon_alpha}, we obtain,

\begin{equation}
\varphi_\en=-\sqrt{6 \,\alpha} \,\arccosh\left(\sqrt[4]{\dfrac{n^2}{12 \alpha }}\right).
\end{equation}

The following step is to calculate the number of e--folding $N$ from Eq. \eqref{Ne} we found that it is given by,

\begin{equation}
\label{Ne_alpha}
N \simeq \dfrac{\sqrt{6\alpha}}{2n}\left(\varphi_\en-\varphi_\ini\right)+\dfrac{3\alpha}{2n}\left[ \sinh\left(\sqrt{\dfrac{2}{3\alpha}}\varphi_\en\right)-\sinh\left(\sqrt{\dfrac{2}{3\alpha}}\varphi_\ini\right)\right],
\end{equation}
where by fixing $N=60$ e--folds we can calculate numerically the value of $\varphi_\ini$ from Eq. \eqref{Ne_alpha}.


In order to obtain the normalized value of the scalar power spectrum, we use the relation of COBE normalization \cite{liddle2000cosmological},

\begin{equation}
\label{normalization}
\delta_R=\dfrac{1}{24 \pi^2}\dfrac{V(\varphi)}{\epsilon(\varphi)},
\end{equation}
by fixing $\delta_R \sim 2.1 \times 10^{-9}$ on the pivot scale $k_*=0.05$ Mpc$^{-1}$.

\section{Background equations}
\label{Background}

The  equations of motion for the scalar field $\varphi$ are given by \cite{liddle2000cosmological},

\begin{eqnarray}
\label{H^2}
H^2 &=& \dfrac{1}{3} \left[V(\varphi)+\dfrac{1}{2}\dot\varphi^2\right],\\
\label{varphi}
\ddot \varphi &+& 3 H \dot{\varphi} = - V_{,\varphi},
\end{eqnarray}
where dots represent the derivation with respect to the physical time $t$, $H$ is the Hubble parameter equal to  $H=\sfrac{\dot{a}}{a}$, being $a$ the scale factor. The term $V(\varphi)$ is the inflationary potential of the scalar field, and $V_{,\varphi}$ is the derivative of the inflationary potential with respect to the scalar field $\varphi$.

\bigskip
\subsection{Numerical solution}
\label{numerical}

\bigskip
The background equations Eqs. \eqref{H^2} and \eqref{varphi} are two coupled differential equations that have to be integrated numerically. To make the numerical integration, we need the initial conditions: $a(0)=1$,  $\varphi(0)=\varphi_\ini$, and $\dot\varphi(0)=\dot\varphi_\sr(0)$, where we use as the initial condition the time derivative of the scalar field $\varphi_\sr(t)$, the solution into the slow--roll approximation.
Using the numerical solution of $\varphi(t)$, we find that the end of inflation occurs when $\varphi(t)=\varphi_\en$ for $t=8.45995 \times 10^7$.
 
In order to apply the phase--integral method we need an analytic expression for the scale factor $a(t)$ and the scalar field $\varphi(t)$, then we made a fit of the numerical solutions for each one.

\bigskip
\subsection{Solution into the Slow--roll approximation}

\bigskip
Into the slow--roll approximation, the inflationary potential $V(\varphi)$ dominates over the kinetic energy, so the background equations Eqs. \eqref{H^2} and   \eqref{varphi} reduce to,

\begin{eqnarray}
\label{H2_alpha}
H^2 &\simeq& \dfrac{1}{3} V(\varphi),\\
\label{varphi_alpha}
3 H \dot\varphi &\simeq& - V_{,\varphi}.
\end{eqnarray}

\bigskip
For the $\alpha$--attractor inflationary model, Eqs. \eqref{H2_alpha} and \eqref{varphi_alpha} have no analytic solution, so we solved them numerically. In order to solve these equations numerically, we need to set the initial conditions: $a(0)=1$, and $\varphi(0)=\varphi_\ini$. From here we obtain the scalar factor $a_\sr(t)$ and  the scalar field $\varphi_\sr(t)$, both into the slow--roll approximation.

\bigskip
\subsubsection{Fit of thes Scale Factor $a(t)$}

\bigskip
The fit of the scale factor $a(t)$ was done using Wolfram Mathematica  until $t_f=4 \times 10^7$ through the following expression \cite{Mathematica:2025},

\bigskip
\begin{equation}
a_\fit(t)=\exp\left\{-\gamma_1 \left[\gamma_2 - \gamma_3 \,t - \gamma_4 \ln\left(\gamma_5 - \gamma_6 \,t\right)\right]\right\},
\label{afit}
\end{equation}
where,

\begin{eqnarray*}
\gamma_1&=&9.5873 \times 10^{-2},\\
\gamma_2&=&21.3366,\\
\gamma_3&=&7.6192\times 10^{-6},\\
\gamma_4&=&6.6741,\\
\gamma_5&=&24.4302,\\
\gamma_6&=&5.2912 \times 10^{-8}.
\end{eqnarray*}

\bigskip
\subsubsection{Fit of the Scalar Field $\varphi(t)$}

\bigskip
The fit of the scalar field $\varphi(t)$ was done using gnuplot   until $t_f=7 \times 10^7$ through the following expression \cite{gnuplot:2024},

\bigskip
\begin{equation}
\varphi_\fit(t)=-\beta_1-\dfrac{\beta_2+\ln\left[\left(2-\dfrac{t}{\beta_3}\right)\beta_4-\beta_5\right]}{\beta_6},
\label{varphifit}
\end{equation}
where

\begin{eqnarray*}
\beta_1&=&1.6435,\\
\beta_2&=&9.2010,\\
\beta_3&=&8.4900\times 10^{7},\\
\beta_4&=&1.4951 \times 10^{-4},\\
\beta_5&=&1.5098\times 10^{-4},\\
\beta_6&=&7.9820.
\end{eqnarray*}

\bigskip
\subsection{Comparison of the plots for the scale factor $a(t)$ and the scalar field $\varphi(t)$ calculated numerically and  with the slow--roll approximation}

\begin{figure}[th!]
\centering
\includegraphics[scale=0.42]{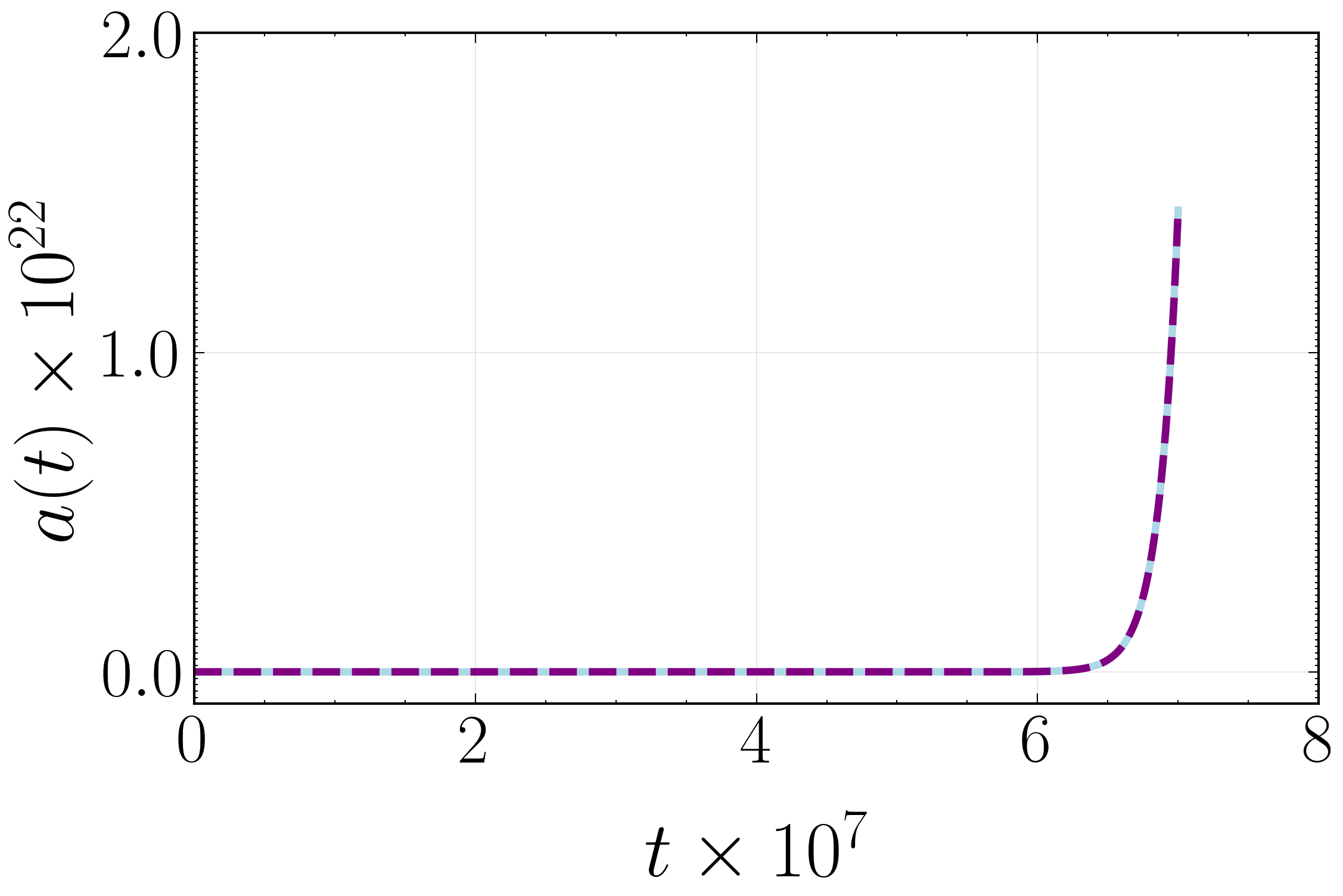}
\caption{Scale factor $a(t)$ for the $\alpha$--attractor inflationary model until $t=7 \times 10^7$. Purple dashed--line represents the numerical solution, and light blue solid  line represents the slow--roll approximation.}
\label{asr_alpha}
\end{figure}

\begin{figure}[th!]
\centering
\includegraphics[scale=0.42]{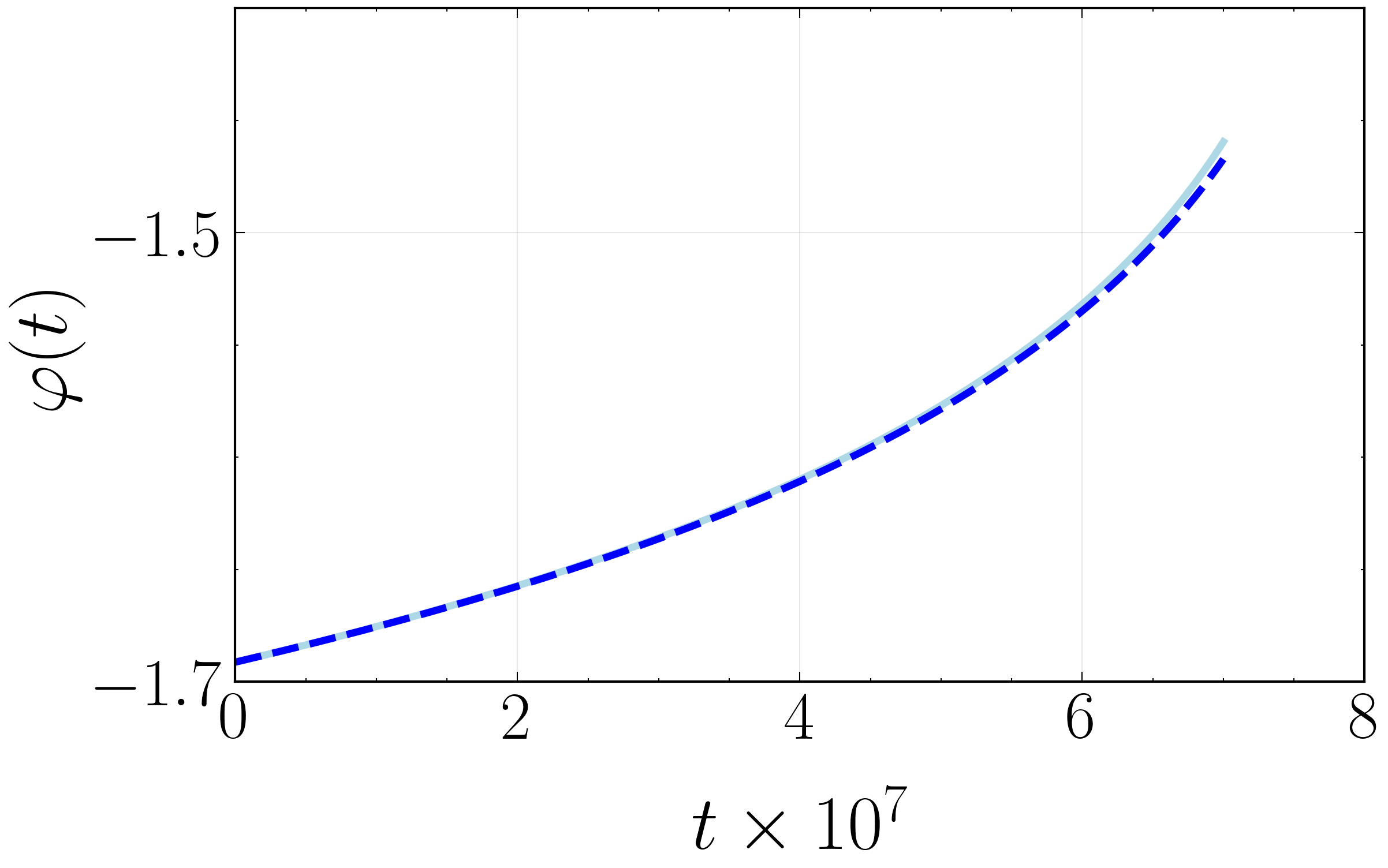}
\caption{Scalar field $\varphi(t)$ for the $\alpha$--attractor inflationary model until $t=7 \times 10^7$. Blue dashed--line represents the numerical solution, and light blue solid represents the slow--roll approximation.}
\label{varphisr_alpha}
\end{figure}

\bigskip
In Figure \ref{asr_alpha}, we can see the plot of the scale factor into the slow--roll approximation $a_\sr(t)$ compared to the scale factor  calculated numerically $a(t)$ for the $\alpha$--attractor inflationary model. We can see that the scale factor has an exponential behavior. In Figure \ref{varphisr_alpha} we can observe the behavior of the scalar field $\varphi(t)$ in both cases.

\bigskip
\subsection{Comparison of the plots of the scale factor $a(t)$ and the scalar field $\varphi(t)$ calculated numerically and with the fit expression}

\begin{figure}[th!]
\centering
\includegraphics[scale=0.42]{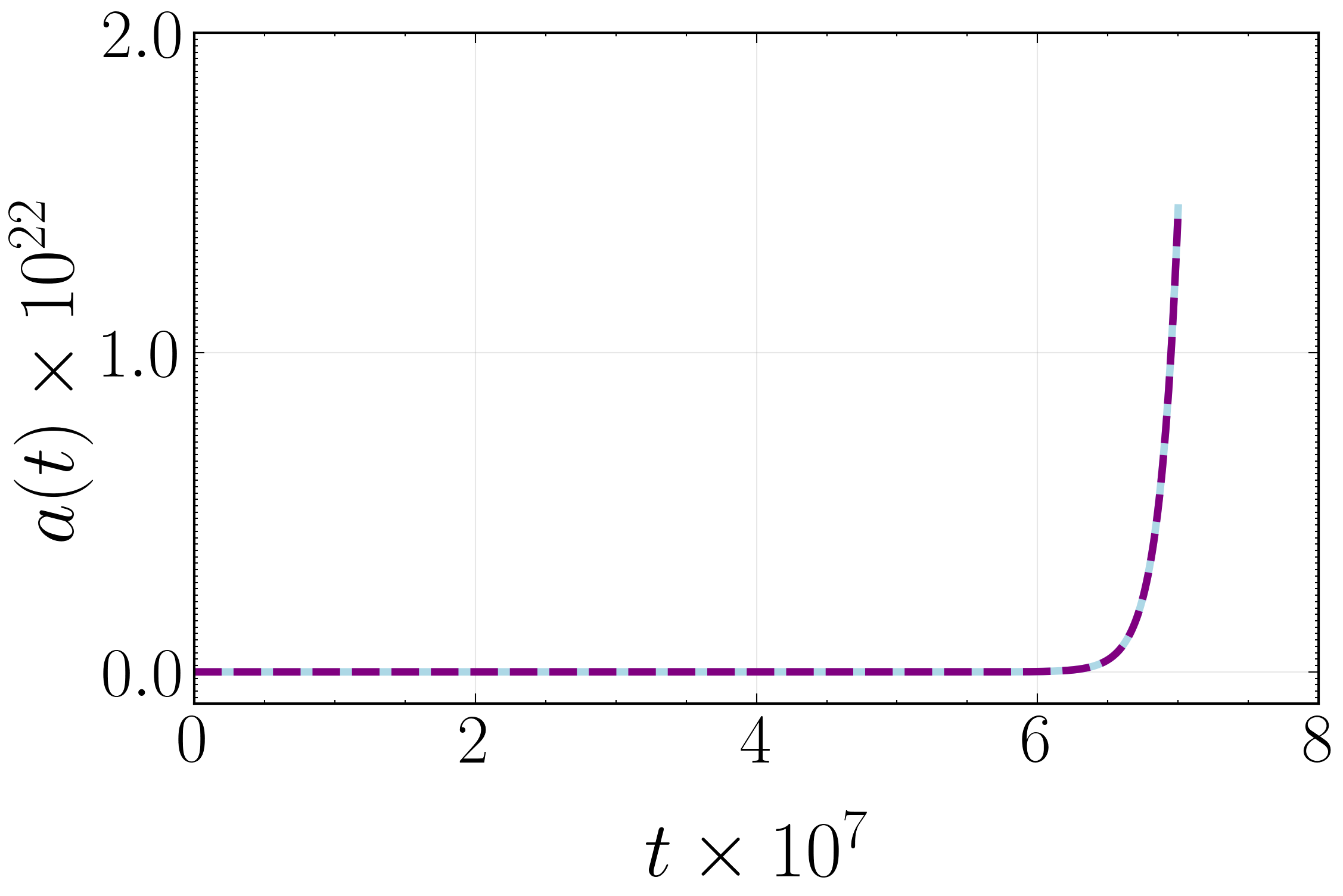}
\caption{Scale factor $a(t)$ for the $\alpha$--attractor inflationary model until $t=7 \times 10^7$. The magenta dashed--line line represents the numerical solution, and  light blue solid line  represents the fit model.}
\label{afit_alpha}
\end{figure}

\begin{figure}[th!]
\centering
\includegraphics[scale=0.42]{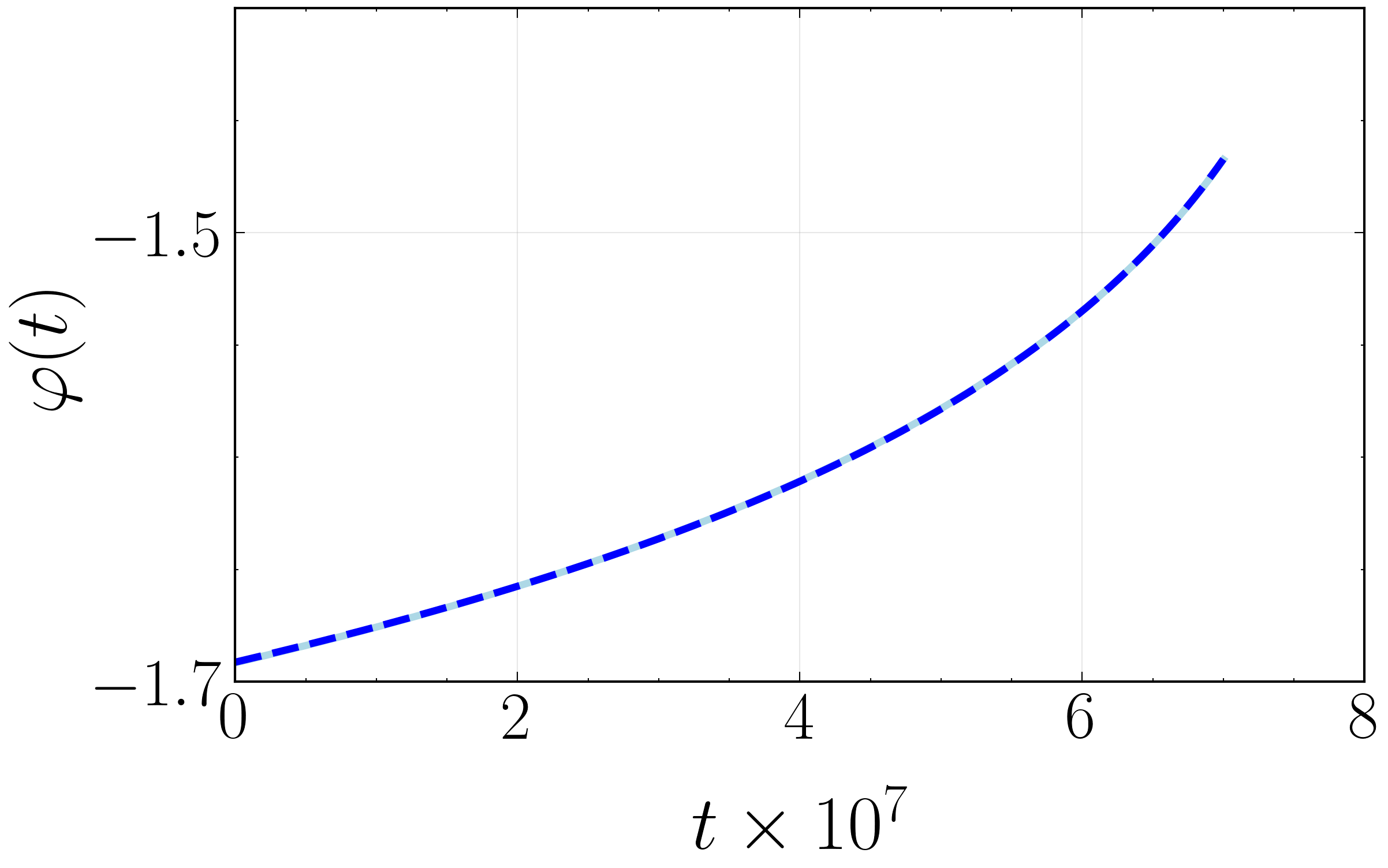}
\caption{Scalar field $\varphi(t)$ for the $\alpha$--attractor inflationary model until $t=7 \times 10^7$. Blue dashed--line line represents the numerical solution, and  light blue  solid line  represents the fit model.}
\label{varphifit_alpha}
\end{figure}

\bigskip
In Figures \ref{afit_alpha} and \ref{varphifit_alpha}, we can see the comparison of the scale factor $a(t)$ and the scalar field $\varphi(t)$  calculated numerically and the fit expression for the $\alpha$--attractor inflationary model. 

\subsection{Statistical comparison of the scale factor $a(t)$ and the scalar field $\varphi(t)$ calculated numerically, with the slow--roll approximation, and with the fit expression}

\bigskip
From Figs. \ref{asr_alpha} -- \ref{varphifit_alpha},  it is difficult to see differences between the numerical results with the slow--roll approximation and the fitted expression. So we calculate the $\chi^2$ statistical factor using,

\begin{equation}
\chi^2 = \sum_{i=1}^{N} \dfrac{\left( A_i - E_i\right)^2}{\sigma_i^2},
\end{equation}
where $A_i$ is the approximate solution, $E_i$ is  the exact numerical solution, $\sigma_i^2$ is the associated uncertainty of each data point. The variance is given by,

\begin{equation}
\sigma_i^2 = \dfrac{1}{N} \sum_{i=1}^{N} (A_i - \bar{A})^2,
\end{equation}
where  $N$ is the number of data points, and $\bar{A}$ is the data mean given by,

\begin{equation}
\bar{A}=\dfrac{1}{N}\sum_{i=1}^{N} A_i.
\end{equation}

In Table \ref{a,varphi}, we present the $\chi^2_i$ statistical factor for the slow--roll  approximation and the fitted expression. From these results, we conclude that the fitted expression provides a better analytical description of the behavior of the scale factor $a(t)$ and the scalar field $\varphi(t)$ than the slow--roll approximation. \textbf{It is important to notice  the relatively high value of $\chi^2_\varphi$, in comparison with $\chi^2_a$,  this is because 
the value of $\varphi(t)$ with each approximate method deviates more from   the numerical result that the scale factor $a(t)$. }\\\\

\begin{table}[th!]
\begin{center}
\caption{$\chi_i^2$ statistical factor  for the scale factor $a(t)$, and the scalar field $\varphi(t)$ calculated with each approximate method respect to the numerical integration.}
\begin{tabular}{c|c|c}
\toprule
Method &   $\chi^2_a$  & $\chi^2_{\varphi}$ \\
\midrule
Slow--roll approximation  &  $7.6381 \times 10^{-3}$& $1.5227 \times 10^4$\\ 
\midrule
Fitted Model &  $5.1755 \times 10^{-4}$& $2.6443 \times 10^3$\\ 
\bottomrule
\end{tabular}
\label{a,varphi}
\end{center}
\end{table}

\section{Equation of perturbations}
\label{Perturbations}

\bigskip
The scalar perturbations are described by the function $u=a\varphi/\varphi'$, where $\varphi$ is a gauge-invariant variable corresponding to the Newtonian potential. The equation of motion of the scalar perturbation $u_k$ in the Fourier space is,

\begin{equation}
\label{dotdotuk}
u_k''+\left(k^2-\dfrac{z_{S}''}{z_{S}}\right)u_k=0,
\end{equation}
where $z_{S}=a\varphi'/\mathcal{H}$, $\mathcal{H}=a'/a$, and the prime indicates the derivative with respect to the conformal time $\eta$. The relation between $t$ and $\eta$ is given via the equation $\D t=a\,\D \eta$.

For tensor perturbations, we introduce the function $v_k=ah$, where $h$ represents the amplitude of the gravitational wave. Tensor perturbations obey a second--order differential equation analogous to Eq. (\ref{dotdotuk}),

\begin{equation}
\label{dotdotvk}
v_k''+\left(k^2-\dfrac{a''}{a}\right)v_k=0.
\end{equation}
Considering the limits  $k^2\gg|z_{S}''/z_{S}|$ (short wavelength) and $k^2\ll|z_{S}''/z_{S}|$  (long wavelength), we have that  the  solutions to
Eq. (\ref{dotdotuk}) exhibit the following asymptotic behavior:

\begin{equation}
\label{boundary_0}
u_k\rightarrow \dfrac{e^{-ik\eta}}{\sqrt{2k}}
\quad \left(k^2\gg|z_{S}''/z_{S}|, -k\eta\rightarrow \infty \right),
\end{equation}

\begin{equation}
\label{boundary_i} u_k\rightarrow A_k z  \quad \left(k^2\ll|z_{S}''/z_{S}|,-k\eta\rightarrow 0\right).
\end{equation}

\noindent Equation \eqref{boundary_0} is used as the initial condition for the perturbations. The same asymptotic conditions hold for tensor
perturbations.

The power spectra for scalar and tensor perturbations are given by the expressions

\begin{eqnarray}
\label{PS}
P_\sca(k)&=& \lim_{kt\rightarrow \infty} \dfrac{k^3}{2 \pi^2}\left|\dfrac{u_k(t)}{z_{S}(t)} \right|^2,\\
\label{PT}
P_\ten(k)&=& \lim_{kt\rightarrow \infty} \dfrac{k^3}{2 \pi^2}\left|\dfrac{v_k(t)}{a(t)} \right|^2,
\end{eqnarray}
from Eq. \eqref{PS} we can calculate the scalar spectral index, defined by,

\begin{equation}
\label{nS}
n_\sca(k)= 1+\dfrac{\D\ln P_\sca(k)}{\D\ln k}.
\end{equation}

\bigskip
\noindent 
In addition, the tensor--to--scalar ratio $r(k)$ is defined as \cite{habib:2005b},

\begin{equation}
\label{R}
r(k)=8\,\dfrac{P_\ten(k)}{P_\sca(k)}.
\end{equation}

In Section~\ref{Background}, we derived the scale factor $a(t)$ and the scalar field $\varphi(t)$ in terms of the physical time $t$, instead of the conformal time $\eta$. Following the same approach, it is necessary to express the equations for the scalar and tensor perturbations, Eqs. \eqref{dotdotuk} and \eqref{dotdotvk}, in terms of the physical time $t$ also. Therefore, the perturbation equations can be written as,

\begin{eqnarray}
\label{dotu}
\ddot{u_k}+\dfrac{\dot{a}}{a}\dot{u_k}+\dfrac{1}{a^2}\left[k^2-{\left(\dot{a}\dot{z_\sca}+a\ddot{z_\sca}\right)a}{z_\sca} \right]u_k&=&0,\\
\label{dotv}
\ddot{v_k}+\dfrac{\dot{a}}{a}\dot{v_k}+\dfrac{1}{a^2}\left[k^2-\left(\dot{a}^2+a\ddot{a}\right) \right]v_k&=&0,
\end{eqnarray}
where $z_\sca$, $\dot{z}_\sca$, and $\ddot{z}_\sca$ are given in terms of $a$, $\varphi$, and their  derivatives with respect to the physical time $t$,

\begin{eqnarray}
\label{zs}
z_\sca&=&\dfrac{a^2 \dot{\varphi}}{\dot{a}},\\
\dot{z}_\sca&=&a\dot{\varphi}\left(2-\dfrac{a \ddot{a}}{\dot{a}^2} \right)+\dfrac{a^2\ddot{\varphi}}{\dot{a}},\\
\nonumber
\ddot{z}_\sca &=& 2\dot{a}\dot{\varphi}+\dfrac{2a^2\ddot{a}^2\dot{\varphi}}{\dot{a}^3}+4a\ddot{\varphi}-\dfrac{a^2\left(2\ddot{a}\ddot{\varphi}+\dddot{a}\dot{\varphi}\right)}{\dot{a}^2}\\
& +& \dfrac{a\left(-2\ddot{a}\dot{\varphi}+a\dddot{\varphi}\right)}{\dot{a}}.
\end{eqnarray}

\bigskip
\section{Solution to the equations of perturbations}
\label{Solution_Perturbations}

\bigskip
\subsection{Numerical Integration}
\label{Numerical_Integration}

\bigskip
The equations for the scalar and tensor perturbations, Eqs. \eqref{dotu} and \eqref{dotv}, are  integrated numerically. Note that $u_k$ and $v_k$ are complex functions, then two differential equations are solved for each of them, one for the real part and  other for the imaginary part. 
Integration is performed in two instances. The first part is done in the limit when $k^2\gg\sfrac{\left(\dot{a}\dot{z_\sca}+a\ddot{z_\sca}\right)a}{z_\sca} $, and $k^2\gg \dot{a}^2+a\ddot{a}$ for scalar and tensor perturbations, respectively, so the differential equations to be solved are the following,

\begin{eqnarray}
\label{dotuk_k2}
\ddot{u_k}&+\dfrac{\dot{a}}{a}\dot{u_k}+\dfrac{k^2}{a^2} u_k=0,\\
\label{dotvk_k2}
\ddot{v_k}&+\dfrac{\dot{a}}{a}\dot{v_k}+\dfrac{k^2}{a^2} v_k=0,
\end{eqnarray}
the integration is done from $t_\ini=0.001\, h_c(k)$ until $t_\en=0.05\, h_c(k)$, where $h_c(k)$ is the horizon defined by the time when the mode freezes $k=a H$. In this stage, we have used Eq. \eqref{boundary_0} as the initial condition  but written in terms of $t$, that is $\tau\simeq -1/[a(t)H]$ since $H$ is nearly constant inside the horizon, then $u_k$ and $v_k$ exhibit oscillatory behavior.

Then, we use the final stage of this solution as an initial condition to solve the complete set of Eqs. \eqref{dotu} and \eqref{dotv} from $k=0.001$\,Mpc$^{-1}$ to $k=10$\,Mpc$^{-1}$, and the integration is done from $t_\ini=0.05\, h_c(k)$ until $t_\en=\, 2 \, h_c(k)$, when the perturbations are already frozen. Finally, from Eqs. \eqref{PS} and \eqref{PT} we calculate the scalar and tensor power spectra. In our calculations, we used the fitted expressions for the scale factor $a_\textnormal{fit}(t)$ given by Eq. \eqref{afit}, and for the scalar field $\varphi_\textnormal{fit}(t)$, given by Eq. \eqref{varphifit}.

Also, in order to find the scalar spectral index $n_\sca(k)$ we implement the fit of the scalar power spectrum $P^\num_\sca(k)$ with a power--law form \cite{giare:2023b,vazquez:2013,das:2023}, that is,

\begin{equation}
\label{PS_fit}
P^\num_\sca(k)=A_\sca\left(\dfrac{k}{k_*}\right)^{n_\sca-1+\dfrac{1}{2}\alpha_\sca\ln\left(\frac{k}{k_*}\right)},
\end{equation}
so that $P^\num_\sca(k)$ becomes scale dependent. Here the symbol `$\num$' means numerical result, $A_\sca$ is the amplitude of the scalar power spectrum, and $\alpha_\sca$ is the running of the scalar spectral index. Equation \eqref{PS_fit} is evaluated on a given pivot scale  $k_*=0.05$\,Mpc$^{-1}$.

Then, to find the tensor--to--scalar ratio $r(k)$ we also utilize the fit of the tensor power spectrum $P^\num_\ten(k)$ with a power--law form \cite{vazquez:2013,vazquez:2020,finelli:2018},

\begin{equation}
\label{PT_fit}
P^\num_\ten(k)= A_\ten\left(\dfrac{k}{k_*}\right)^{n_\ten+\dfrac{1}{2}\alpha_\ten\ln\left(\frac{k}{k_*}\right)},
\end{equation}
where $A_\ten$ is the amplitude of the tensor power spectrum, and $\alpha_\ten$ is the running of the tensor spectral index. Equation \eqref{PT_fit} is evaluated at a given pivot scale $k_*=0.002$\,Mpc$^{-1}$.

The relation between the scalar power spectrum $P_\sca(k)$ and the amplitude of the scalar power spectrum $A_\sca$, without considering the running of the scalar spectral index $\alpha_\sca$, is given by \cite{akrami:2020},

\begin{equation}
\label{AS}
\ln P_\sca(k)=\ln A_\sca+\left(n_\sca-1\right)\ln\left(\dfrac{k}{k_*}\right),
\end{equation}
here, Planck takes as a pivot scale $k_{*}=0.05$\,Mpc$^{-1}$ \cite{akrami:2020}. Finally, we evaluate the scalar power spectrum $P_\sca(k)$ at $k=0.05$\,Mpc$^{-1}$, therefore $P_\sca(k)$ becomes equal to $A_\sca$.


\bigskip
\subsection{Slow--roll approximation}
\label{slow--roll}

\bigskip
We are interested in computing the scalar and tensor power spectra for the $\alpha$--attractor inflation model into the slow--roll approximation, for that, we use the following expressions \cite{adshead:2011,stewart:1993},

\begin{eqnarray}
\label{PS_sr}
P^\sr_\sca(k)&\simeq&\dfrac{1}{12\pi^2}
\left[1+2\, C_\sca \left(2 \,\epsilon+\eta\right)-2 \epsilon\right]\dfrac{V^3}{V'^2},\\
\label{PT_sr}
P^\sr_\ten(k)&\simeq&\dfrac{1}{12\pi^2}\left(1- 2\, C_\ten\, \epsilon \right) V,
\end{eqnarray}
where $C_\sca=2-\ln 2 - b \simeq 0.729637$, here $b$ is the Euler--Mascheroni constant, and
$C_\ten=\ln 2 + b - 1 \simeq = 0.2704$. The symbol  `$\sr$' means the slow--roll approximation.


\bigskip
Furthermore, the observables in the slow--roll approximation are given by Eqs. \eqref{nS_initial} and \eqref{r_initial},

\begin{eqnarray}
\label{nS_sr}
n^\sr_\sca(k) &\simeq& 1- 4 \,\epsilon -2\, \eta,\\
\label{r_sr}
r^\sr(k) &\simeq& 16 \,\epsilon,
\end{eqnarray}
where $\eta$ and $\epsilon$ are given by Eqs. \eqref{epsilon} and \eqref{eta}, respectively.

\bigskip
\subsection{Uniform approximation method}

\textbf{This method is discussed in the Appendix.}

\bigskip
\subsection{Phase--integral method} 

\textbf{This method is discussed in the Appendix.}

\section{Solution to the Perturbation Equations}
\label{Methods}

\subsection{Numerical Integration}

\bigskip
After solving numerically the scalar and tensor perturbation equations, we fitted the scalar power spectrum $P^\mathrm{numfit}_\sca(k)$ and the tensor power spectrum $P^\mathrm{numfit}_\ten(k)$ using Eqs. \eqref{PS_fit} and \eqref{PT_fit}. It is important to note that these perturbation equations were derived using the fitted analytical expressions for $a_\fit(t)$ and $\varphi_\fit(t)$, so we used the symbol `$\mathrm{numfit}$' to identify these solutions. In this way the scalar power spectrum is given by,

\begin{equation}
\label{PS_fitnum}
P^\mathrm{numfit}_\sca(k)=A_\sca\left(\dfrac{k}{k_*}\right)^{n_\sca-1+\dfrac{1}{2}\alpha_\sca\ln\left(\frac{k}{k_*}\right)},
\end{equation}
where  $A_\sca=2.2219 \times 10^{-9}$, $n_\sca=0.9607$, $\alpha_\sca=-7.9933 \times 10^{-4}$, evaluated at  the pivot scale $k_*=0.05$\,Mpc$^{-1}$. Similarly, for the tensor power spectrum we have,

\begin{equation}
\label{PT_numfit}
P^\mathrm{numfit}_\ten(k)= A_\ten\left(\dfrac{k}{k_*}\right)^{n_\ten+\dfrac{1}{2}\alpha_\ten\ln\left(\frac{k}{k_*}\right)},
\end{equation}
where $A_\ten=1.3462 \times 10^{-14}$, $n_\ten=-6.7899 \times 10^{-6}$, $\alpha_\ten=-3.5201 \times 10^{-7}$, and the pivot scale  is $k_*=0.002$\,Mpc$^{-1}$.

\bigskip
\subsection{Slow--roll Approximation}
\bigskip

Using Eqs. \eqref{PS_sr}--\eqref{r_sr} we obtain the following expressions for the scalar and tensor power spectra together with the observables into the slow--roll approximation for the $\alpha$--attractors inflationary model,

\begin{eqnarray}
\nonumber
P^\sr_{\sca}(k)&=&\dfrac{\lambda}{12 n^2 \pi^2} e^{-n\tanh\left(\dfrac{\varphi_*}{\sqrt{6\alpha}}\right)}
 \Bigg\{
6\alpha\, \cosh^4\left(\dfrac{\varphi_*}{\sqrt{6\alpha}}\right) \\
&+&  n\bigg[(4\, C_\textnormal{S} - 1) \, n + 2\, C_\textnormal{S}\, \sinh\left(\sqrt{\dfrac{2}{3 \alpha}}\varphi_*\right)\bigg]\Bigg\},\\
\nonumber
P^\sr_{\ten}(k)&=&\dfrac{\lambda}{72\, \pi^2\, \alpha}
e^{-n \tanh\left( \dfrac{\varphi_*}{\sqrt{6\alpha}} \right)}  \left[
6\alpha - C_\textnormal{T}\, n^2\, \operatorname{sech}^4\left( \dfrac{\varphi_*}{\sqrt{6\alpha}} \right)\right],\\\\
n^\sr_{\sca}(k)&=&
1 - \dfrac{n}{ 6\alpha}\, \operatorname{sech}^4 \left( \dfrac{\varphi_*}{\sqrt{6\alpha}} \right) \left[ n - 2 \,\sinh \left( \sqrt{\dfrac{2}{3\alpha}}\, \varphi_*  \right) \right],\\
r^\sr(k)&=&
\dfrac{4 n^2}{3 \alpha}\, \operatorname{sech}^4\left( \dfrac{\varphi_*}{\sqrt{6\alpha}} \right),
\end{eqnarray}
where $\varphi_*$ is calculated at the horizon for each case.

\subsection{Uniform Approximation Method}

\bigskip
The equations governing the scalar and tensor perturbations, Eqs. \eqref{alpha_ddotUk} and \eqref{alpha_ddotVk}, are solved using the uniform approximation method described in Subsection \ref{uniform}. We choose:

\begin{eqnarray}
r_\sca(k,t)&=&R_\sca(k,t),\\ 
r_\ten(k,t)&=&R_\ten(k,t),
\end{eqnarray}
which satisfies the validity condition \eqref{validity_uniform}. 

Using the analytical expressions for the scale factor $a_\textnormal{fit}(t)$ and the scalar field $\varphi_\textnormal{fit}(t)$ we obtain that the functions $R_\sca(k,t)$, and $R_\ten(k,t)$ for the $\alpha$--attractors inflationary model are given by,

\begin{eqnarray}
\label{RS_uniform}
\nonumber
R_\sca(k,t) &=& \dfrac{1}{4 \left(\gamma_5 - \gamma_6\, t \right)^2} \Bigg\{4 \,e^{2 \gamma_1 (\gamma_2 - \gamma_3 t )} k^2 \left(\gamma_5 - \gamma_6 \,t \right)^{2(1 -  \gamma_1 \gamma_4)} \\
\nonumber
&+& \dfrac{8 \,\gamma_3 \gamma_4 \gamma_6^3 \left( -\gamma_5 + \gamma_6  \,t \right)}{\left[- \gamma_3 \gamma_5 + (\gamma_4 + \gamma_3\, t  )\gamma_6 \right]^2}- 9 \,\gamma_1^2 \left[ -\gamma_3 \gamma_5 + (\gamma_4 + \gamma_3 \, t)\gamma_6 \right]^2 \\
\nonumber
&-& \dfrac{8 \left( \gamma_5 -  \gamma_6 \, t \right)^2 \beta_4^2}{\left[ - 2 \,\beta_3 \beta_4 + \beta_4 \, t + \beta_3 \beta_5 \right]^2} + \dfrac{8\, \gamma_4 \gamma_6^2 \left( \gamma_5 -  \gamma_6 \,t\right) \beta_4}{
\left[ \gamma_3 \gamma_5 - ( \gamma_4 +  \gamma_3 \,t)\,\gamma_6 \right]\left(-2 \,\beta_3 \beta_4 + \beta_4 \,t + \beta_3 \beta_5 \right)} \\
\nonumber
&+& \dfrac{6\, \gamma_1 \left\{
2 \gamma_3 (\gamma_5 - \gamma_6 \, t)^2 \beta_4 + 
\gamma_4 \gamma_6 \left[ 
-2 \,\gamma_5 \beta_4 + \gamma_6 \left( 2 \,\beta_3 \beta_4 + \beta_4 \,t- \beta_3 \beta_5 \right)\right]\right\}}{
-2\, \beta_3 \beta_4 +\beta_4 \,t + \beta_3 \beta_5}\Bigg\},
\\\\
\nonumber
\label{RT_uniform}
R_\ten(k,t)&=&
e^{2 \gamma_1 (\gamma_2 -  \gamma_3\, t)}\, k^2 \left(\gamma_5 -  \gamma_6\, t\right)^{-2 \gamma_1 \gamma_4}\\
&+& {3\, \gamma_1 \bigg[2 \gamma_4 \gamma_6^2 - 3 \gamma_1 \left(-\gamma_3 \gamma_5 +  \gamma_3 \gamma_6 \, t + \gamma_4 \gamma_6\right)^2\bigg]}{4 \left(\gamma_5 - \gamma_6 \, t\right)^2}.
\end{eqnarray}

\bigskip
Using Eqs. \eqref{RS_uniform} and \eqref{RT_uniform}, we first calculate the functions $U_k(k, t)$ and $V_k(k, t)$, as well as $\rho_{\textnormal{S},\textnormal{T}}(k, t)$. These quantities allow us to determine $u_k^\ua(t)$ and $v_k^\ua(t)$, which are then used to compute the scalar and tensor power spectra for the $\alpha$--attractors inflationary model via the uniform approximation method, as described by Eqs. \eqref{PS_ua} and \eqref{PT_ua}.

\bigskip
\subsection{Phase--Integral Method}

\bigskip
In order to solve Eq. \eqref{alpha_ddotUk} and  Eq. \eqref{alpha_ddotVk}   with the help of the phase--integral   method, we choose the following base functions $Q_{\sca,\ten}(k,t)$ for scalar and tensor perturbations,

\begin{eqnarray}
\label{Q1}
Q_\sca^2(k,t)&=&R_\sca(k,t),\\
\label{Q2}
Q_\ten^2(k,t)&=&R_\ten(k,t),
\end{eqnarray}
where  $R_\sca(k,t)$ and  $R_\ten(k,t)$ are given by  Eq. \eqref{RS_uniform} and \eqref{RT_uniform}, respectively.  Using this selection, the phase--integral   method is  valid as  $k t\rightarrow \infty$, limit where we should impose the condition \eqref{alpha_cero_Uk}, where the validity condition  $\mu \ll 1$, given by Eq. \eqref{validity_phase},  holds. The selection, given in Eq. (\ref{Q1}) and Eq. \eqref{Q2}, makes the first order phase--integral  method coincide with the WKB solution. The basis functions  $Q^2_\sca(k,t)$ and  $Q^2_\ten(k,t)$ have turning points  $t_\ret=\tau_\sca=1.47004 \times 10^7$ for the mode $k=0.05\,\Mpc^{-1}$ and $t_\ret=\tau_\ten=1.03032 \times 10^7$ for the mode $k=0.002\,\Mpc^{-1}$, where the turning point represents the horizon.

With the help of the analytic expressions for the scale factor $a_\textnormal{fit}(t)$ and the scalar field $\varphi_\textnormal{fit}(t)$,  we calculated the functions $q_\sca(k,t)$,  $q_\ten(k,t)$, $\omega_\sca(k,t)$, and $\omega_\ten(k,t)$ in order to compute the functions $u_k^{\phai}(t)$ and $v_k^{\phai}(t)$.

Finally,   by applying Eqs. \eqref{PS_pi} and \eqref{PT_pi}, the scalar and tensor power spectra are obtained using the phase--integral method up to third order in approximation, for the $\alpha$--attractors inflationary model.

\section{Results}
\label{Results}

\begin{figure}[th!]
\centering
\includegraphics[scale=0.5]{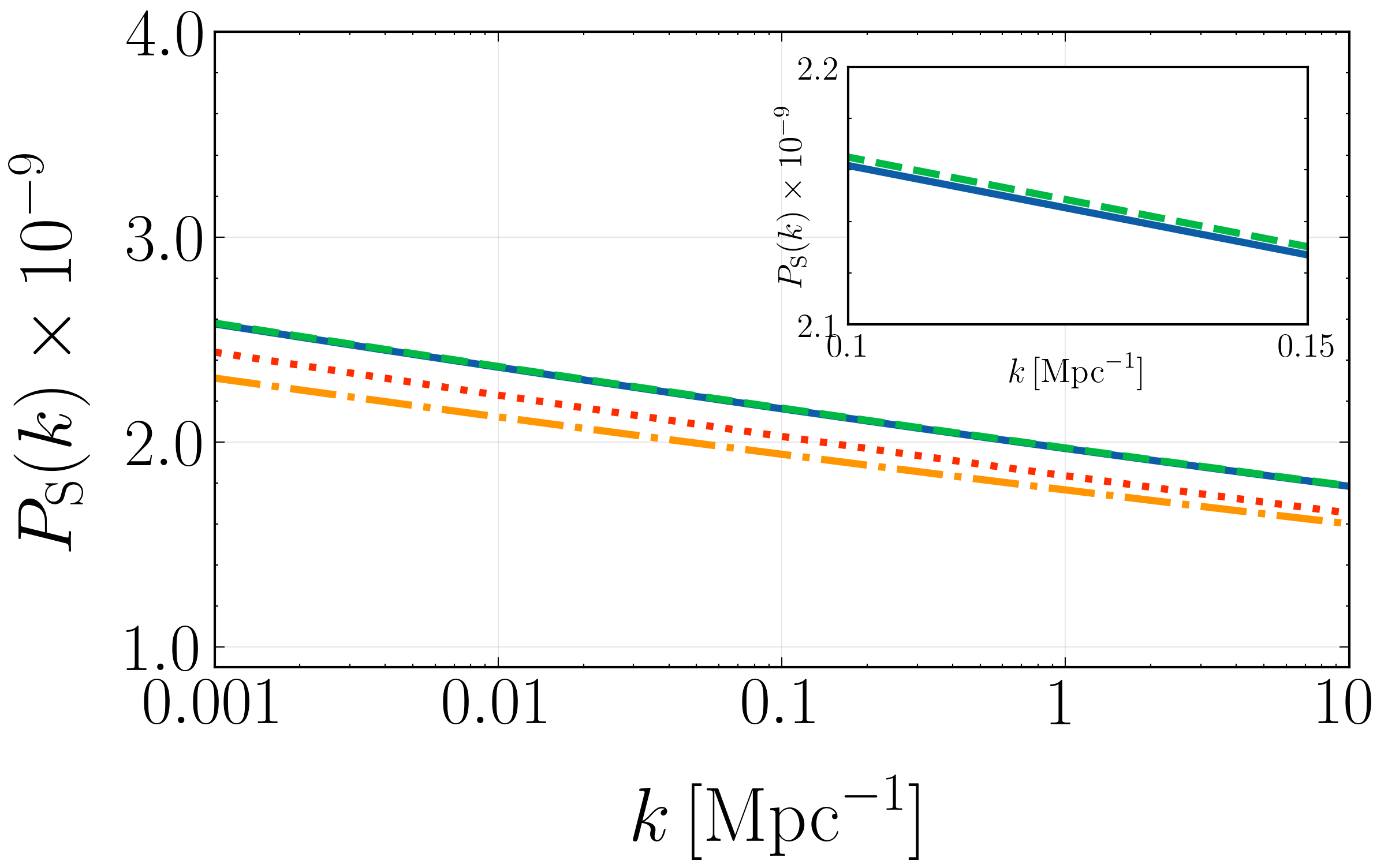}
\caption{Scalar power spectrum $P_\sca(k)$  for the $\alpha$--attractor inflationary model.  Solid blue line: numerical result; light green dashed line: third--order phase--integral approximation; orange dot--dashed line: uniform approximation method; red dotted line: slow--roll approximation.  The inset is an enlargement of the figure.}
\label{PS_methods}
\end{figure}

\begin{figure}[th!]
\centering
\includegraphics[scale=0.5]{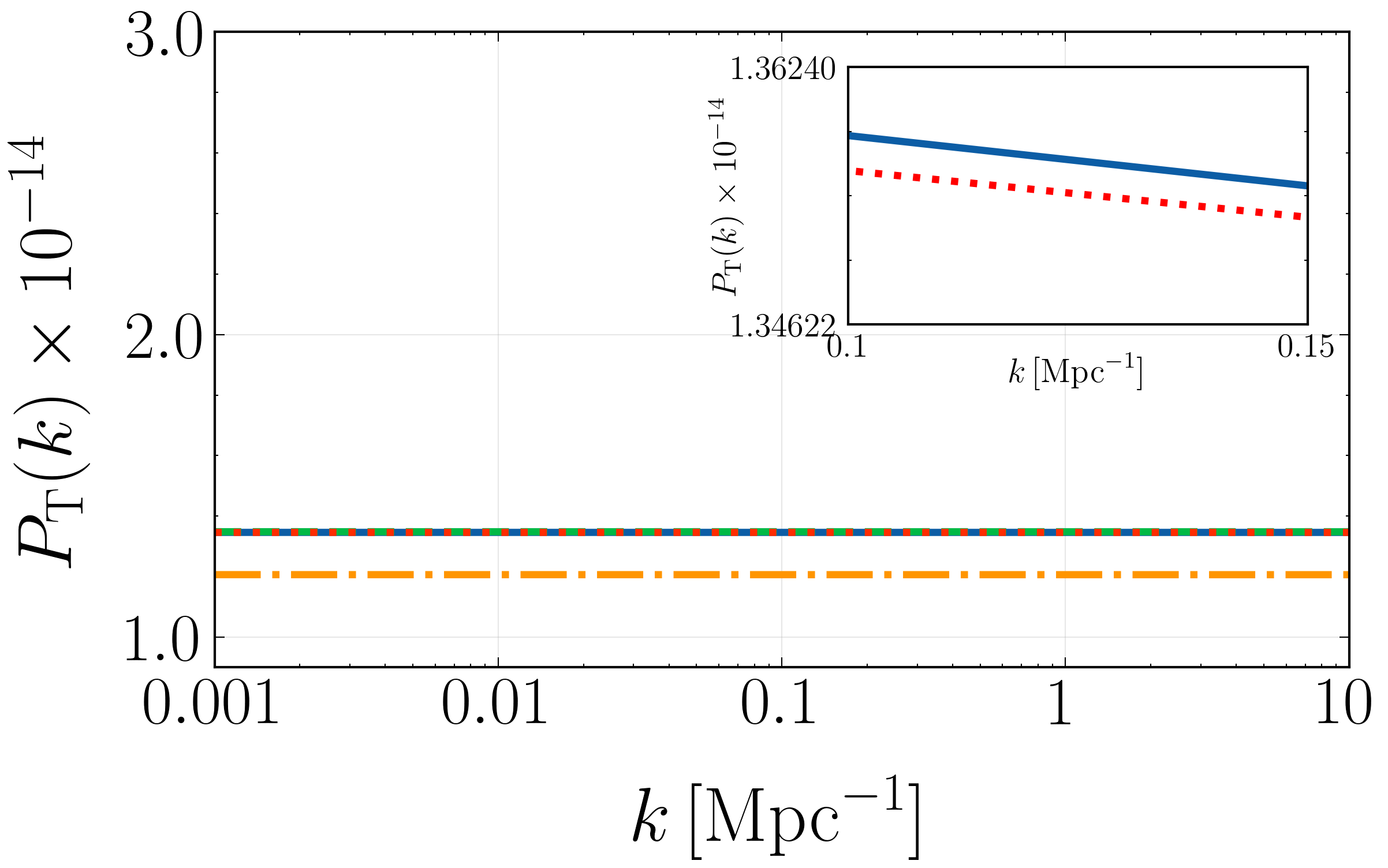}
\caption{Tensor power spectrum $P_\ten(k)$  for the $\alpha$--attractor inflationary model.  Solid blue line: numerical result; light green dashed line: third--order phase--integral approximation; orange dot-dashed line: uniform approximation method; red dotted line: slow--roll approximation.  The inset is an enlargement of the figure.}
\label{PT_methods}
\end{figure}

In Figure \ref{PS_methods} we can observe the scalar power spectrum calculated: a) numerically, b) using the third--order phase--integral method, c) using the uniform approximation method, and d) using the slow--roll approximation as we indicated in Section \ref{Methods}. From the inset of this Figure, we can say that the third--order phase integral method is the best result of our three approximation methods, which can be confirmed with the calculation of $\chi^2_{P_\sca}$ shown in Table \ref{chi2_PST}. In Figure \ref{PT_methods} we can observe the tensor power spectrum calculated: a) numerically, b) using the third--order phase--integral method, c) using the uniform approximation method, and d) using the slow--roll approximation as we indicated in Section \ref{Methods}. From the inset of this Figure, we can say that the slow--roll approximation is the best result of our three approximation methods, which can be confirmed with the calculation of $\chi^2_{P_\ten}$ shown in Table \ref{chi2_PST}.

\begin{table}[th!]
\begin{center}
\caption{$\chi_i^2$ statistical factor for the scalar power spectrum $P_\sca(k)$, and the tensor power spectrum $P_\ten(k)$ of each approximate method respect to the numerical integration.}
\begin{tabular}{c|c|c}
\toprule
Method &   $\chi^2_{P_\sca}$  & $\chi^2_{P_\ten}$ \\
\midrule
Slow--roll Approximation  &  $3.5192 \times 10^3$& $3.6138\times 10^3$\\ 
Uniform Approximation  &  $1.1374 \times 10^4$& $1.3768 \times 10^{11}$\\
Phase--integral  &  $1.9751$& $3.8254 \times 10^8$\\ 
\bottomrule
\end{tabular}
\label{chi2_PST}
\end{center}
\end{table}

We also computed the observables for the $\alpha$--attractor inflationary model, which are summarized in Tables  \ref{observables_pi3}, \ref{observables_ua}, \ref{observables_sr}.
For each approximate method, we present the results along with a comparison to the numerical solution, expressed in terms of the percentage relative error. The comparison is performed at the pivot scales $k_*=0.05\,\text{Mpc}^{-1}$ for the scalar quantities, and $k_*=0.002\,\text{Mpc}^{-1}$ for the tensor quantities.

\begin{table}[th!]
\caption{Observables for the  $\alpha$--attractor inflationary model calculated: $a)$ numerically and $b)$ using the  third--order phase--integral method, both evaluate at its corresponding  pivot scale $k_*$.}
\begin{center}
\resizebox{\textwidth}{!}{
\begin{tabular}{cccc}
\toprule
Observables                                           & Numerical & $3^{\mathrm{rd}}$--order phase--integral method&rel. err. (\%) \\  
\midrule 
$\ln\left(10^{10} A_{\sca_{0.05}} \right) $   & $\;\;3.1001$   & $\;\;3.1025$& $7.6772 \times 10^{-2}$\\
$n_{\sca_{0.05}}$                                            & $\;\;0.9607$ &$\;\;0.9606 $&  $6.2454 \times 10^{-3}$\\
$r_{0.002} $                                     & $\;\;4.2883 \times 10^{-5}$ & $\;\; 4.2882 \times 10^{-5}$&$2.3319 \times 10^{-3}$ \\
\bottomrule
\end{tabular}
}
\label{observables_pi3}
\end{center}
\end{table}

\begin{table}[th!]
\caption{Observables for the   $\alpha$--attractor inflationary model calculated: $a)$ numerically and $b)$ using the  uniform  approximation method, both evaluate at its corresponding  pivot scale $k_*$.}
\begin{center}
\resizebox{\textwidth}{!}{
\begin{tabular}{cccc}
\toprule
Observables                                           & Numerical & Uniform approximation method &rel. err. (\%) \\  
\midrule 
$\ln\left(10^{10} A_{\sca_{0.05}} \right) $   & $\;\;3.1001$   & $\;\; 2.9927$& $3.4644$\\
$n_{\sca_{0.05}}$                                            & $\;\;0.9607$ &$\;\; 0.96065 $&  $  5.2045\times 10^{-3}$\\
$r_{0.002} $                                     & $\;\; 4.2883 \times 10^{-5}$ & $\;\;  4.2827 \times 10^{-5}$&$ 1.3058 \times 10^{-1}$ \\
\bottomrule
\end{tabular}
}
\label{observables_ua}
\end{center}
\end{table}

\begin{table}[th!]
\caption{Observables for the   $\alpha$--attractor inflationary model calculated: $a)$ numerically and $b)$ using the  slow--roll approximation, both evaluate at its corresponding  pivot scale $k_*$.}
\vspace{-0.5cm}
\begin{center}
\resizebox{\textwidth}{!}{
\begin{tabular}{cccc}
\toprule
Observables                                          & Numerical & Slow--roll approximation&rel. err (\%) \\  
\midrule 
$\ln\left(10^{10} A_{\sca_{0.05}} \right) $   & $\;\;3.1001$   & $\;\;3.0381$& $1.999$\\
$n_{\sca_{0.05}}$                                            & $\;\;0.9607$ &$\;\;0.9584 $&  $2.3940 \times 10^{-1}$\\
$r_{0.002} $                                     & $\;\;4.2883 \times 10^{-5}$ & $\;\; 4.5363 \times 10^{-5}$&$5.7832$ \\
\bottomrule
\end{tabular}
}
\label{observables_sr}
\end{center}
\end{table}

In table \ref{observables_comparation}, we can see the values for the scalar spectral index $n_\sca(k)$ and the tensor-to-scalar ratio $r(k)$ for the $\alpha$--attractors inflationary model and the Starobinsky inflationary model, both evaluated at their corresponding pivot scale. 

\begin{table}[th!]
\caption{Observables of the $\alpha$--attractor inflationary model  and the Starobinsky inflationary model calculated: 
$a)$ numerically, $b)$ using the third--order phase integral approximation, $c)$ using  the uniform approximation method, and $d)$ the slow--roll approximation,  evaluated at 
their corresponding pivot scale $k_*$.}
\begin{center}
\resizebox{\textwidth}{!}{
\begin{tabular}{ccccc}
\toprule
Method & Observables & $\alpha$--attractor & Starobinsky \\  
\bottomrule 
\multirow{2}{*}{Numerical} 
& $n_{\sca_{0.05}}$ & $\;\;0.9607$ & $\;\;0.9682$ \\
& $r_{0.002}$       & $\;\;4.2883 \times 10^{-5}$ & $\;\ 2.6337 \times 10^{-3}$ \\
\midrule
\multirow{2}{*}{$3^{\mathrm{rd}}$--order phase--integral method} 
& $n_{\sca_{0.05}}$ & $\;\;0.9606$ & $\;\;0.9683$ \\ 
& $r_{0.002}$       & $\;\;4.2883 \times 10^{-5}$ & $\;\;2.6338 \times 10^{-3}$ \\
\midrule
\multirow{2}{*}{Uniform approximation method} 
& $n_{\sca_{0.05}}$ & $\;\;0.9607$ & $\;\;0.9683$\\
& $r_{0.002}$       & $\;\;4.2827\times 10^{-5}$ & $\;\;2.6310\times 10^{-3}$ \\
\midrule
\multirow{2}{*}{Slow--roll approximation} 
& $n_{\sca_{0.05}}$ &  $\;\;0.9584$ & $\;\;0.9659$ \\
& $r_{0.002}$       & $\;\;4.5363 \times 10^{-5}$ & $\;\;2.9815 \times 10^{-3}$\\
\bottomrule
\end{tabular}
}
\label{observables_comparation}
\end{center}
\end{table}

\begin{figure}[th!]
\centering
\includegraphics[scale=0.65]{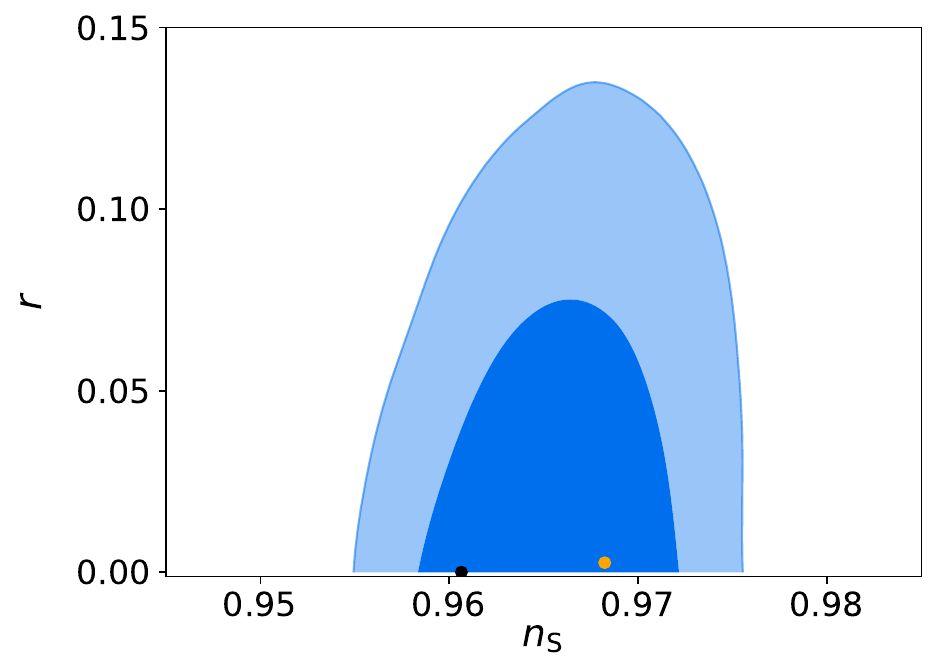}
\caption{The contour plot $(n_\sca,r)$ for the $\alpha$--attractor inflationary model (black point) and the Starobinsky inflationary model (yellow point).}
\label{contour}
\end{figure}

In Table \ref{observables_error} we can also see the percentage relative error for the scalar spectral index $n_\sca(k)$ compared with the values reported by Planck 2018 results where $n_\sca=0.9659$,and $r<0.10$ ($95 \%$ CL, \textit{Planck} TT + lowE + lensing) \cite{akrami:2020}. From this Table, we can conclude that although the scalar spectral index $n_\sca$ for the Starobinsky model presents lower relative error with respect to the observational results, the values of $r$ for the $\alpha$--attractor inflationary model give a lower value for the tensor--to-scalar ratio. In Figure \ref{contour} we can observe the contour plot $(n_\sca,r)$ for both models calculated numerically; from this plot, we can conclude that both models are a good representation of the inflationary epoch because both lie within the $95 \%$ confidence region.

\begin{table}[th!]
\caption{Percentage relative error for the observables of the $\alpha$--attractor inflationary model  and the Starobinsky inflationary model calculated: 
$a)$ numerically, $b)$ using the third--order phase integral approximation, $c)$ using  the uniform approximation method, and $(d)$ the slow--roll approximation,  evaluated at 
their corresponding pivot scale $k_*$.}
\begin{center}
\resizebox{\textwidth}{!}{
\begin{tabular}{ccccc}
\midrule
Method & Observable & $\alpha$--attractor & Starobinsky \\ 
&        & rel. err. (\%) &  rel. err. (\%)\\
\bottomrule
Numerical
& $n_{\sca_{0.05}}$ & $0.53836$ & $0.2381$ \\
$3^{\mathrm{rd}}$--order phase--integral method
& $n_{\sca_{0.05}}$ &  $0.54457$ & $0.2485$\\ 
Uniform approximation method
& $n_{\sca_{0.05}}$ &  $0.54354$ & $0.2485$\\
Slow--roll approximation 
& $n_{\sca_{0.05}}$ & $0.77648$ &  $0.001035$\\
\bottomrule
\end{tabular}
}
\label{observables_error}
\end{center}
\end{table}

\newpage
\section{Conclusions}
\label{Conclusions}

In this work, we  solved numerically the scalar and tensor perturbation equations for the $\alpha$--attractor inflationary model using the slow--roll approximation, the uniform approximation method, and the third--order phase--integral method. This inflationary model allows us to describe both the initial accelerated expansion of the Universe during the inflationary epoch and the current accelerated expansion with the same potential. Once the scalar power spectrum $P_\sca(k)$, the tensor power spectrum $P_\ten(k)$, and the observables are calculated, we find that the third--order phase--integral method provides excellent agreement with the numerical results, providing more accurate results than both the slow--roll and uniform approximation methods,\textbf{ at least for the scalar power spectrum. However the slow--roll approximation performs best results for the tensor 
power spectrum. Which can be deduced from the  comparison of the power spectra obtained from each approximation method with the numerical results, for which we employ the statistical factor $\chi_i^2$. Also we compare the observables of the $\alpha$--attractor inflationary model calculated with each approximation method to those obtained from the numerical solution, using the percentage relative error, obtained better results with those calculated with the $3^\textnormal{rd}$--order phase--integral--method. Additionally we compare our model with the Starobinsky inflationary model and we can conclude that both models are a good representation of the inflationary epoch because both lie within the $95 \%$ confidence region.}

\section{Appendix}
\label{Appendix}
\subsection{Uniform approximation method}
\label{uniform}

\bigskip
In order to apply semiclassical methods, we eliminate the terms $\dot{u}_k$ and $\dot{v}_k$ in Eq. \eqref{dotu} and Eqs. \eqref{dotv}. We make the change of variables $u_k(t)=\sfrac{U_k(t)}{\sqrt{a}}$ and $v_k(t)=\sfrac{V_k(t)}{\sqrt{a}}$, obtaining that $U_{k}$ and $V_{k}$ satisfy the differential equations

\begin{eqnarray}
\label{alpha_ddotUk}
\ddot{U}_k+R_\sca(k,t)U_k&=&0,\\
\label{alpha_ddotVk}
\ddot{V}_k+R_\ten(k,t)V_k&=&0,
\end{eqnarray}
with,

\begin{eqnarray}
\label{RS}
R_\sca(k,t)&=&\dfrac{1}{a^2}\left[k^2-\dfrac{\left(\dot{a}\dot{z_\sca}+a\ddot{z_\sca}\right)a}{z_\sca} \right]+\dfrac{1}{4a^2}\left(a^2-2a\ddot{a}\right),\\
\label{RT}	
R_\ten(k,t)&=&\dfrac{1}{a^2}\left[k^2-\left(\dot{a}^2+a\ddot{a}\right) \right]+\dfrac{1}{4a^2}\left(a^2-2a\ddot{a}\right),
\end{eqnarray}
where $U(k)$ satisfies the asymptotic conditions

\begin{eqnarray}
\label{alpha_cero_Uk}
U_k&\rightarrow&A_k \sqrt{a(t)} \,z_\sca(t),\quad  k\,t\rightarrow \infty,\\
\label{alpha_borde_Uk}
U_k&\rightarrow&\sqrt{\dfrac{a(t)}{2k}}\exp{\left[-i\, k\, \eta(t)\right]}, \quad k\,t\rightarrow 0,
\end{eqnarray}
the asymptotic conditions (\ref{alpha_cero_Uk}) and (\ref{alpha_borde_Uk}) also hold for $V_k$.

We want to obtain an approximate solution of the differential equation \eqref{alpha_ddotUk} and  \eqref{alpha_ddotVk}  in terms of the known solutions $w_\sca(\rho_\sca)$ and $w_\ten(\rho_\ten)$ of the comparison equation \cite{berry:1972,habib:2002,rojas:2007b,rojas:2007c,dingle}: 

\begin{eqnarray}
\label{comparison_Sca}
\dfrac{\D ^2 w_\sca(\rho_\sca)}{\D \rho_\sca^2}+r_\sca(\rho_\sca) w_\sca(\rho_\sca)=0,\\
\label{comparison_Ten}
\dfrac{\D ^2 w_\ten(\rho_\ten)}{\D \rho_\ten^2}+r_\ten(\rho_\ten) w_\ten(\rho_\ten)=0,
\end{eqnarray}
where $r_\sca(\rho_\sca)$ is chosen similar to $R_\sca(k, t)$ and $r_\ten(\rho_\ten)$ is chosen similar to $R_\ten(k, t)$,  with the same number of zeros, so that the solutions of equations \eqref{comparison_Sca} and \eqref{comparison_Ten} are known. 

The functions $U(k,t)$ and $V(k,t)$ must also be similar to $w_\sca(\rho_\sca)$ and $w_\ten(\rho_\ten)$, they can be related via \cite{berry:1972},

\begin{eqnarray}
\label{relationS}
U_k(k,t) \approx \left\{\dfrac{r_\sca\left[\rho_\sca(k,t)\right]}{R_\sca(k,t)} \right\}^{\sfrac{1}{4}} w_\sca\left[\rho_\sca(k,t)\right],\\
\label{relationT}
V_k(k,t) \approx \left\{\dfrac{r_\ten\left[\rho_\ten(k,t)\right]}{R_\ten(k,t)} \right\}^{\sfrac{1}{4}} w_\ten\left[\rho_\ten(k,t)\right].
\end{eqnarray}

The validity condition to Eqs. \eqref{relationS} and \eqref{relationT} being a good solution is given by

\begin{equation}
\label{validity_uniform}
\left|\dfrac{1}{R_{\sca,\ten}(t)} \left(\dfrac{\D \rho_{\sca,\ten}}{\D t} \right)\dfrac{\D^2}{\D t^2}\left(\dfrac{\D \rho_{\sca,\ten}}{\D t} \right)^{\sfrac{1}{2}}  \right|\ll 1.
\end{equation}

Eqs. \eqref{relationS} and \eqref{relationT} give a uniform approximation for $U_k(k,t)$ and $V_k(k,t)$ for the entire range of $t$, including the turning points.

\begin{figure}[htbp]
\begin{center}
\subfigure[]{
\label{alpha_QSa}
\includegraphics[scale=0.4]{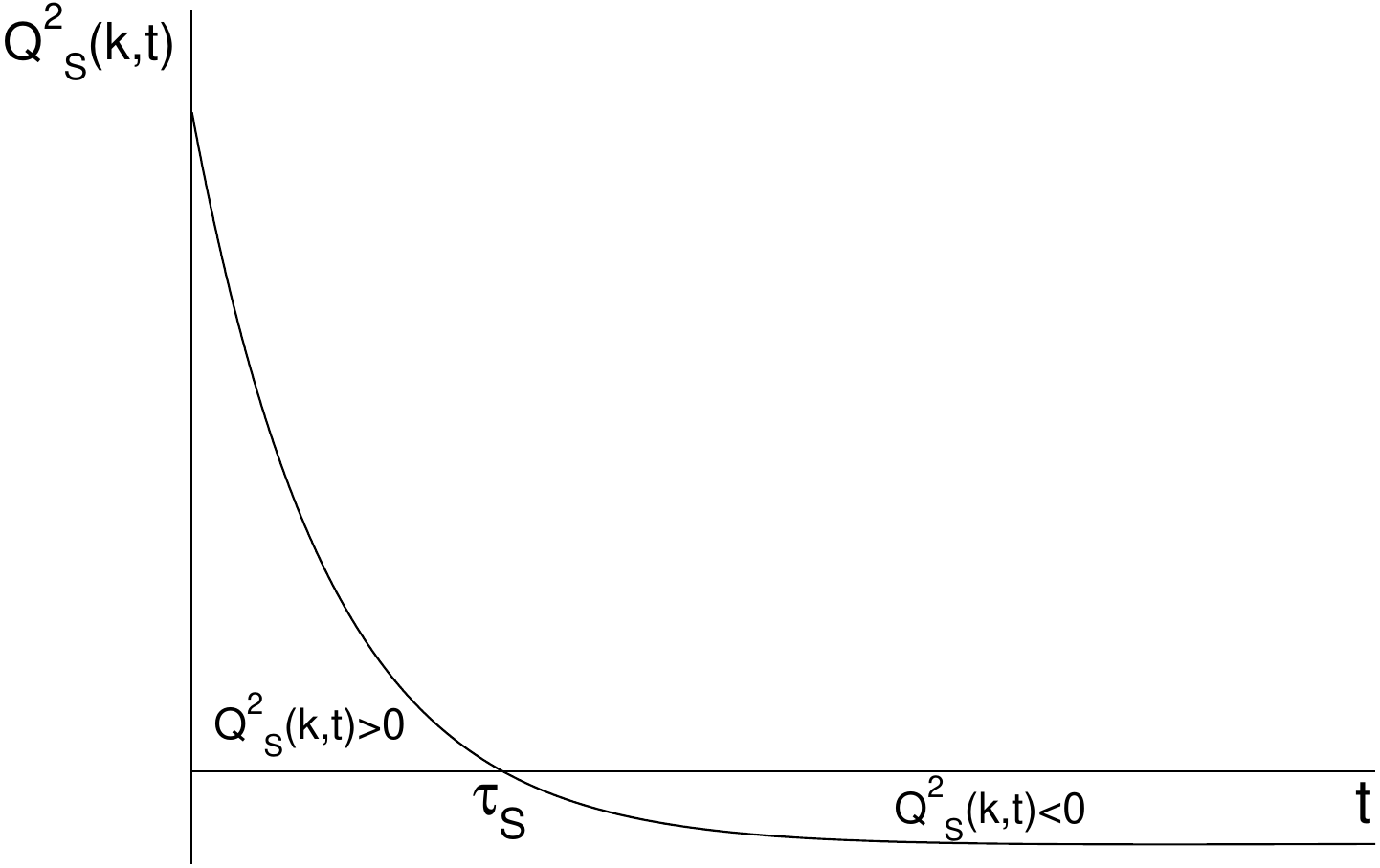}}
\subfigure[]{
\label{alpha_QSb}
\includegraphics[scale=0.35]{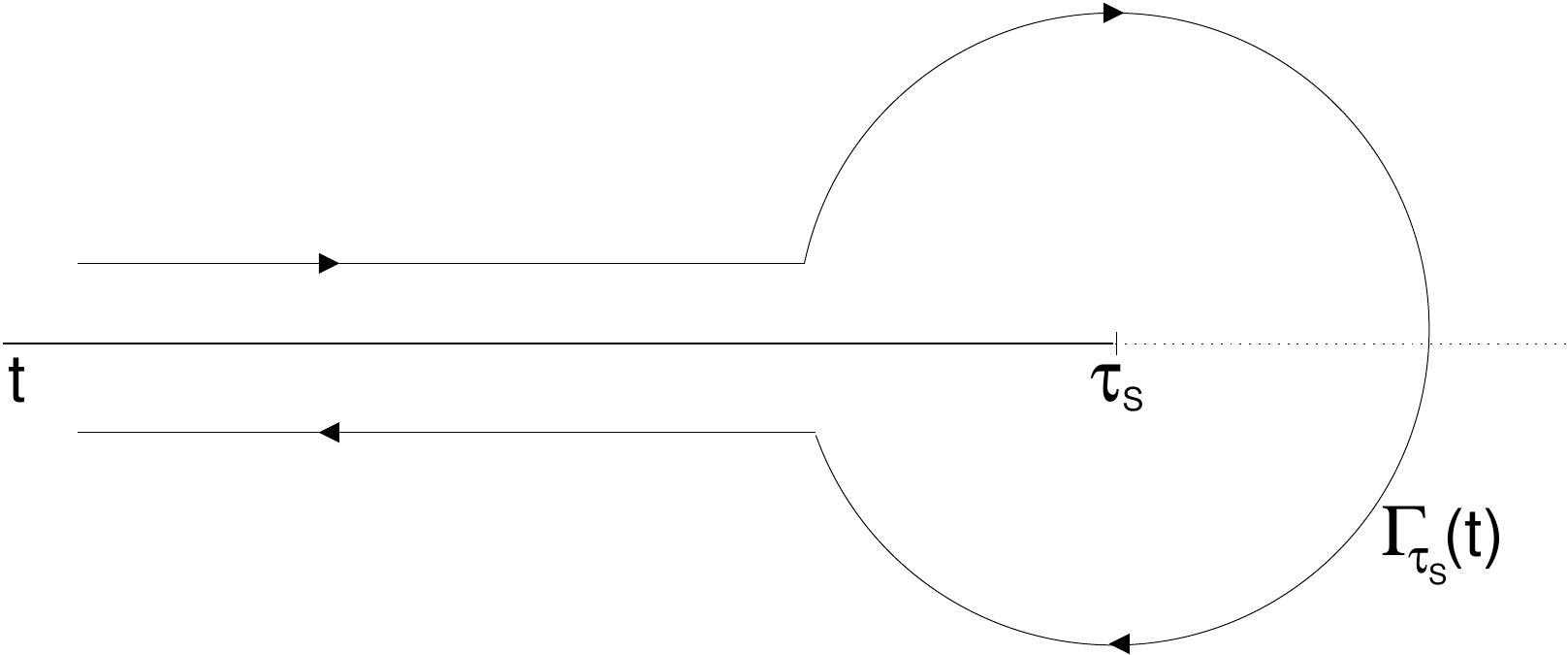}}
\subfigure[]{
\label{alpha_QSc}
\includegraphics[scale=0.35]{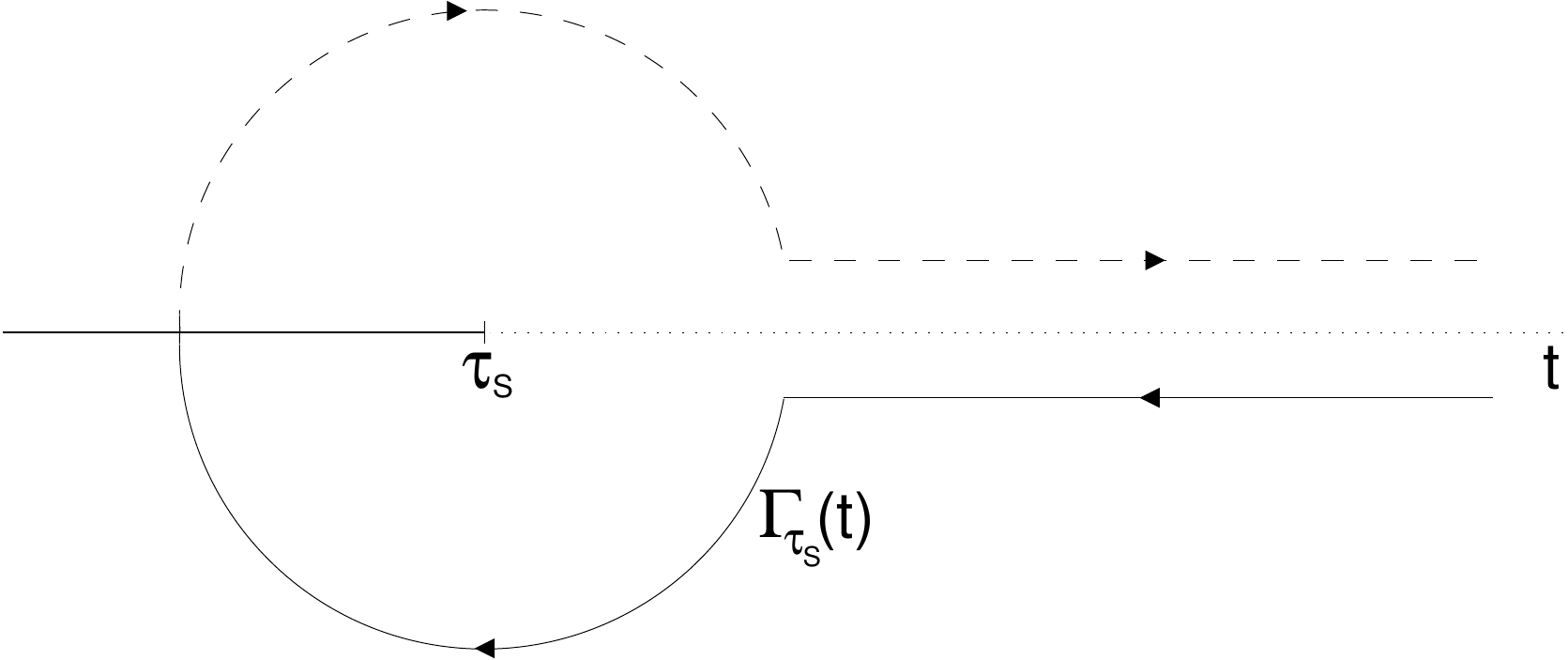}}
\caption{\small{(a) Behavior of the function  $Q_\sca^2(k,t)$. (b) Contour of integration $\Gamma_{\tau_\sca}(t)$ for  $0<t<\tau_\sca$. (c) Contour of  integration  $\Gamma_{\tau_\sca}(t)$ for $t>\tau_\sca$. The dashed line indicates the part of the path on the second Riemann sheet.}}
\end{center}
\end{figure}

\begin{figure}[htbp]
\subfigure[]{
\label{alpha_QTa}
\includegraphics[scale=0.4]{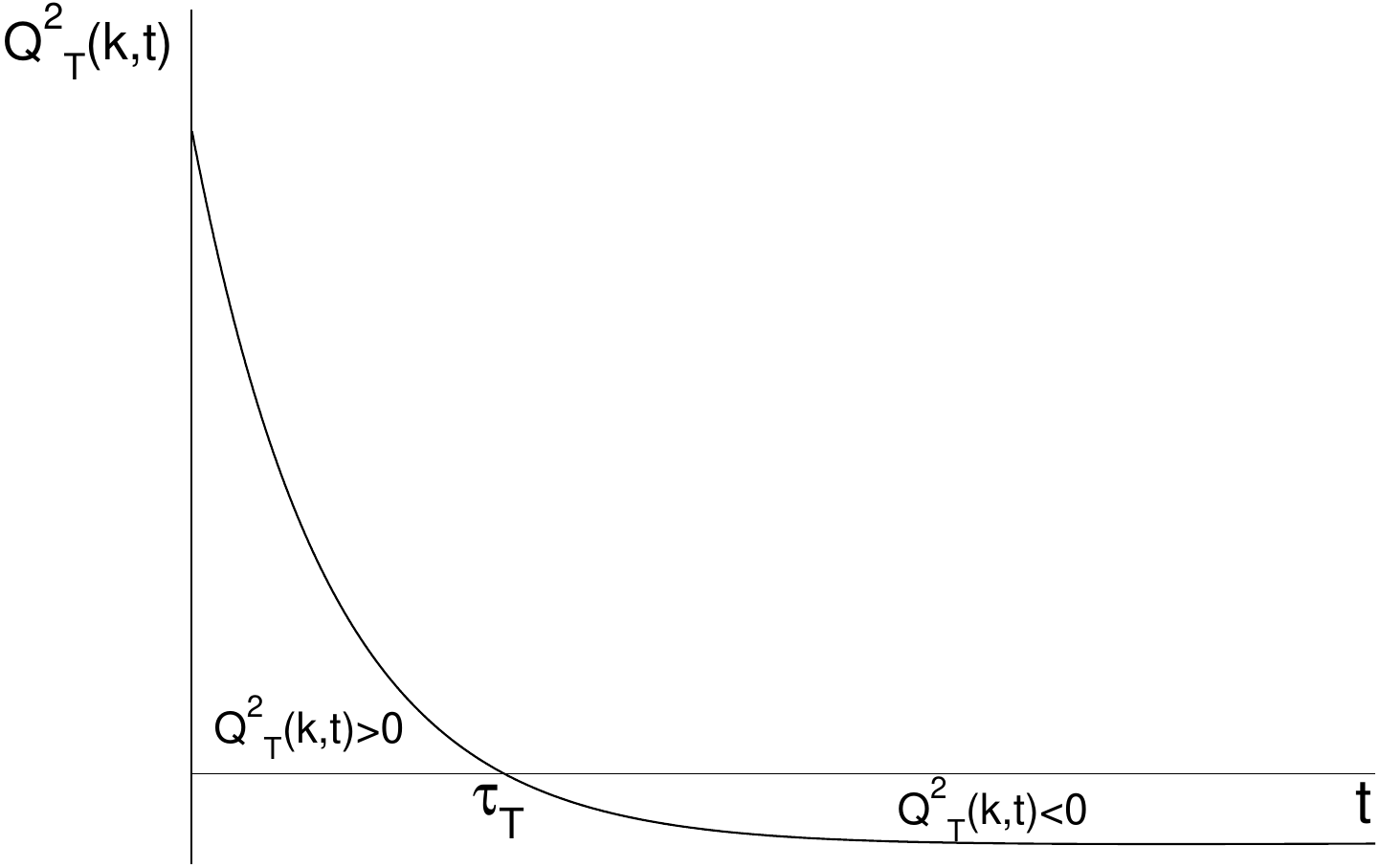}}
\subfigure[]{
\label{alpha_QTb}
\includegraphics[scale=0.35]{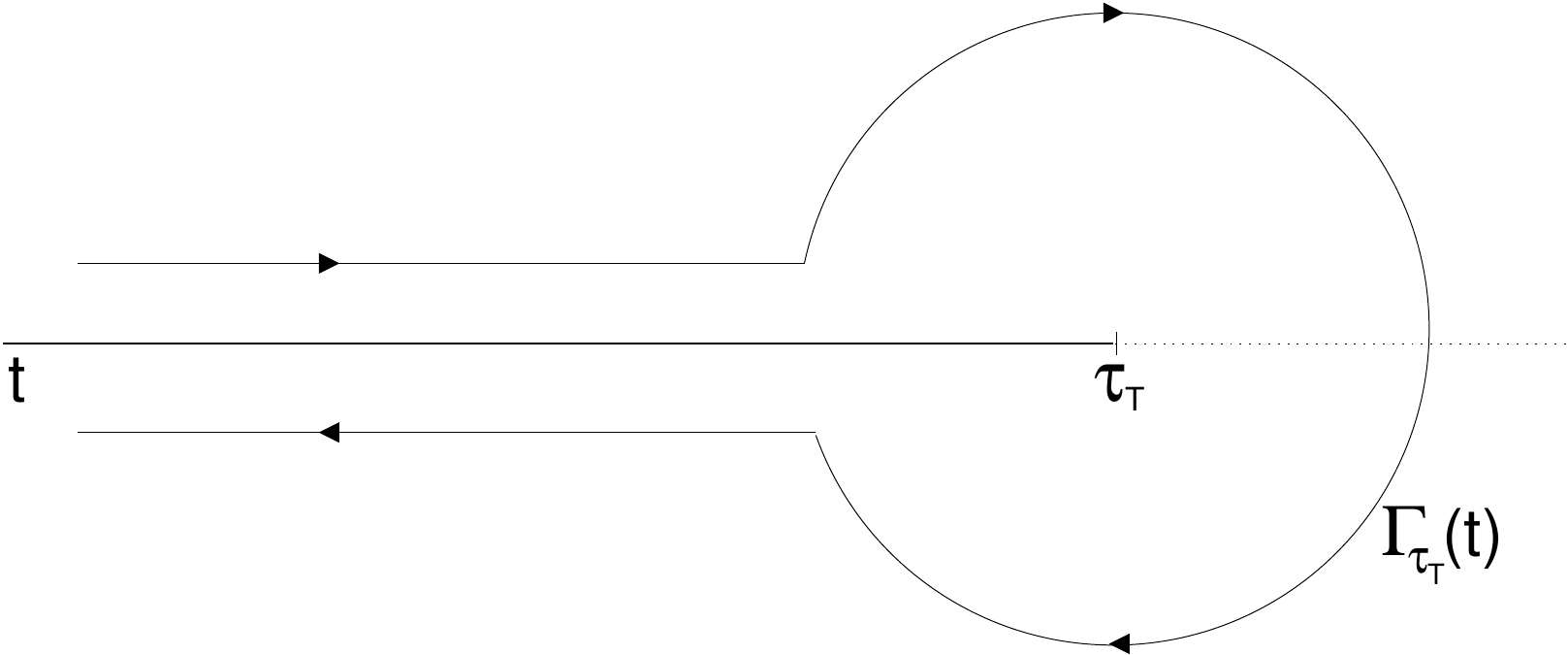}}
\subfigure[]{
\label{alpha_QTc}
\includegraphics[scale=0.35]{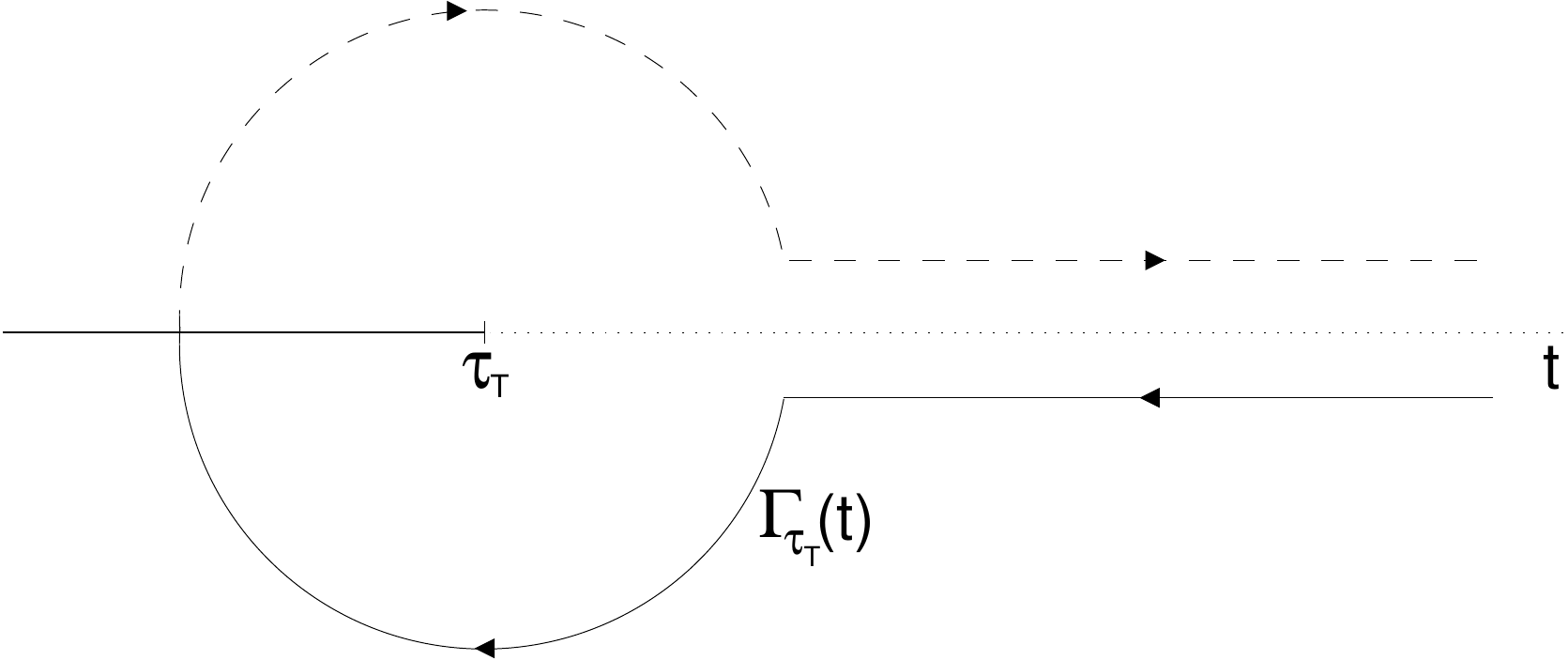}}
\caption{\small{(a) Behavior of  $Q_\ten^2(k,t)$.
(b) Contour of integration $\Gamma_{\tau_\ten}(t)$ for $0<t<\tau_\ten$. 
(c) Contour of integration  $\Gamma_{\tau_\ten}(t)$ for $t>\tau_\ten$. 
The dashed lined indicates the part of the path on the second Riemann sheet.}}
\end{figure}

Find an approximate solution to the differential equations \eqref{alpha_ddotUk} and \eqref{alpha_ddotVk} in a region where  $Q_\sca^2(k,t)$ and   $Q_\ten^2(k,t)$ have a simple root in $t_\ret=\tau_\sca$, and $t_\ret=\tau_\ten$, respectively, so that $Q_{\sca,\ten}^2(k,t)>0$ for $0<t<t_\ret$ and $Q_{\sca,\ten}^2(k,t)<0$ for  $t>t_\ret$ as depicted in Figure \ref{alpha_QSa} and Figure \ref{alpha_QTa}.

A suitable comparison function is $ r_\sca(\rho)=\pm\, \rho_\sca  $ and $r_\ten(\rho)=\pm\, \rho_\ten$, therefore there are two cases: 

\begin{description}
\item{a)} In the classically allowed region, $Q_{\sca,\ten}^2(k, t)> 0$,  we choose $r_{\sca,\ten}(\rho_{\sca,\ten}) = + \,\rho_{\sca,\ten}$ and the comparison equations to solved are 
	
\begin{eqnarray}
\label{eqr,rk=rho_Sca}
\dfrac{\D ^2w_\sca}{\D \rho_\sca^2}+\rho_\sca\, w_\sca=0,\\
\label{eqr,rk=rho_Ten}
\dfrac{\D ^2w_\ten}{\D \rho_\ten^2}+\rho_\ten w_\ten=0.
\end{eqnarray}
	
Eq. \eqref{eqr,rk=rho_Sca}  and \eqref{eqr,rk=rho_Ten} are the Airy equation that has two independent solutions $A_i(-\rho_{\sca,\ten}) $ and $B_i(-\rho_{\sca,\ten})$ \cite{abramowitz:1965}. The mapping relation is given by \cite{berry:1972} 
	
\begin{equation}
\label{mapeo,rk=rho}
\dfrac{\D \rho_{\sca,\ten}}{\D t}=\left[{Q_{\sca,\ten}^2(k,t)}{\rho_{\sca,\ten}}\right]^{\sfrac{1}{2}}.
\end{equation}
	
Finally, the approximate solutions to the differential equations \eqref{alpha_ddotUk} and \eqref{alpha_ddotVk}  are

\begin{eqnarray}
\label{Uk_zero}
\nonumber
U_k(k,t)&=&\left[\dfrac{\rho^\ele_{\sca}(k,t)}{Q_\sca^2(k,t)} \right]^{\sfrac{1}{4}} \left\{C_1
A_i[-\rho^\ele_{\sca}(k,t)]+C_2 B_i[-\rho^\ele_{\sca}(k,t)] \right\},\\
\\
\label{Vk_zero}
\nonumber
V_k(k,t)&=&\left[\dfrac{\rho^\ele_{\ten}(k,t)}{Q_\ten^2(k,t)} \right]^{\sfrac{1}{4}} \left\{C_1
A_i[-\rho^\ele_{\ten}(k,t)]+C_2 B_i[-\rho^\ele_{\ten}(k,t)] \right\},\\
\\
\dfrac{2}{3}\left[\rho^\ele_{\sca,\ten}(k,t)\right]^{\sfrac{3}{2}}&=&\int_{t}^{t_\ret} \left[Q_{\sca,\ten}^2(k,t)\right]^{\sfrac{1}{2}}\D t,
\end{eqnarray}
\medskip
where  $C_1$ and  $C_2$ are two constants to be determined with the help of the boundary conditions \eqref{alpha_borde_Uk}.  
In the limit $kt \rightarrow \infty$, the asymptotic formulas are used  \cite{abramowitz:1965}

\begin{eqnarray}
\label{Ai_1}
A_i(-\rho)&\sim&\pi^{-\sfrac{1}{2}}\rho^{-\sfrac{1}{4}}\sin\left(\dfrac{2}{3}\rho^{\sfrac{3}{2}}+\dfrac{\pi}{4}\right),\\
B_i(-\rho)&\sim&\pi^{-\sfrac{1}{2}}\rho^{-\sfrac{1}{4}}\cos\left(\dfrac{2}{3}\rho^{\sfrac{3}{2}}+\dfrac{\pi}{4}\right).
\label{Bi_1}
\end{eqnarray}

It is found that $C_1=\sqrt{\dfrac{\pi}{2}}\e^{-i\sfrac{\pi}{4}}$ and $C_2=\sqrt{\dfrac{\pi}{2}}\e^{i\sfrac{\pi}{4}}$.

\item{b)} In the classically forbidden region, $Q_{\sca,\ten}^2(k, t)< 0$, we choose $r_{\sca,\ten}(\rho_{\sca,\ten}) = - \,\rho_{\sca,\ten}$,    and solve the comparison equations

\begin{eqnarray}
\label{eqr,rk=rho_Sca_right}
\dfrac{\D ^2w_\sca}{\D \rho_\sca^2}-\rho_\sca\, w_\sca=0,\\
\label{eqr,rk=rho_Ten_right}
\dfrac{\D ^2w_\ten}{\D \rho_\ten^2}-\rho_\ten\, w_\ten=0.
\end{eqnarray}

Eq. \eqref{eqr,rk=rho_Sca_right}  and \eqref{eqr,rk=rho_Ten_right} has the form of the Airy differential equation, which has two independent solutions $A_i(-\rho_{\sca,\ten}) $ and $B_i(-\rho_{\sca,\ten}) $ \cite{abramowitz:1965}. The mapping relation is given by \cite{berry:1972} 

\begin{equation}
\label{mapeo,rk=rho_right}
\dfrac{\D \rho_{\sca,\ten}}{\D t}=\left[{-Q_{\sca,\ten}^2(k,t)}{\rho_{\sca,\ten}}\right]^{1/2}.
\end{equation}

The approximate solutions to the differential equations \eqref{alpha_ddotUk} and \eqref{alpha_ddotVk}  are

\begin{eqnarray}
\label{Uk_infinity}	
\nonumber
U_k(k,t)&=&\left[{-\rho^\ere_\sca(k,t)}{Q_\sca^2(k,t)} \right]^{1/4} \left\{C_1
A_i[\rho^\ere_\sca(k,t)]+C_2 B_i[\rho^\ere_\sca(k,t)] \right\},\\
\\
\label{Vk_infinity}
\nonumber
V_k(k,t)&=&\left[{-\rho^\ere_\ten(k,t)}{Q_\ten^2(k,t)} \right]^{1/4} \left\{C_1
A_i[\rho^\ere_\ten(k,t)]+C_2 B_i[\rho^\ere_\ten(k,t)] \right\},\\
\\
\frac{2}{3}\left[\rho^\ere_{\sca,\ten}(k,t)\right]^{3/2}&=&\int_{t_\ret}^{t} \left[-Q_{\sca,\ten}^2(k,t)\right]^{1/2}\D t.
\end{eqnarray}

For the computation of the power spectrum we need to take the limit $k\,t\rightarrow \infty$ of the solutions  \eqref{Uk_infinity} and \eqref{Vk_infinity}. In this limit we have

\begin{eqnarray}
\label{Ai_right}
A_i(\rho)&\sim&\dfrac{\pi^{-\sfrac{1}{2}}}{2}\rho^{-\sfrac{1}{4}}\exp\left(-\dfrac{2}{3}\rho^{\sfrac{3}{2}}\right),\\
B_i(\rho)&\sim&\pi^{-\sfrac{1}{2}}\rho^{-\sfrac{1}{4}}\exp\left(\dfrac{2}{3}\rho^{\sfrac{3}{2}}\right).
\label{Bi_right}
\end{eqnarray}

Finally,

\begin{eqnarray}
\label{limit_uk}
\nonumber
u_k^\ua(t)&\rightarrow&  \dfrac{C}{\sqrt{2\,a(t)}}\left[-Q_\sca^2(k,t)\right]^{-\sfrac{1}{2}}\left\{ \dfrac{1}{2}\exp\left(-\int_{\tau_\sca}^{t}\left[-Q_\sca^2(k,t)\right]^{\sfrac{1}{2}} \D t\right)\right.\\
& +&\left.\im\,\exp\left(\int_{\tau_\sca}^{t}\left[-Q_\sca^2(k,t)\right]^{\sfrac{1}{2}} \D t\right)\right\},\\
\nonumber
\label{limit_vk}
v_k^\ua(t)&\rightarrow&  \dfrac{C}{\sqrt{2\,a(t)}}\left[-Q_\ten^2(k,t)\right]^{-\sfrac{1}{2}}\left\{ \dfrac{1}{2}\exp\left(-\int_{\tau_\ten}^{t}\left[-Q_\ten^2(k,t)\right]^{\sfrac{1}{2}} \D t\right)\right.\\
& +&\left.\im\,\exp\left(\int_{\tau_\ten}^{t}\left[-Q_\ten^2(k,t)\right]^{\sfrac{1}{2}} \D t\right)\right\},
\end{eqnarray}
where $C$ is a phase factor.  Using the growing part of the solutions   \eqref{limit_uk} and  \eqref{limit_vk}   one can compute the scalar and tensor power spectra using the uniform approximation method,

\end{description}
\begin{eqnarray}	
\label{PS_ua}
P^\ua_\sca(k)&=&\lim_{-k t\rightarrow \infty} \dfrac{k^3}{2\pi^2} \left|\dfrac{u_k^{\ua}(t)}{z_\sca(t)}\right|^2,\\
\label{PT_ua}
P^\ua_\ten(k)&=&\lim_{-k t\rightarrow \infty} \dfrac{k^3}{2\pi^2} \left|\dfrac{v_k^\ua(t)}{a(t)}\right|^2,
\end{eqnarray} 
where `$\ua$' means the uniform approximation method.


\bigskip
\subsection{Phase--integral method} 
\label{phase}
\bigskip

Let us consider the differential equation,

\begin{equation}
\label{eq_ori}
\dfrac{\D^2 u_k}{\D z^2}+R(z) \,u_k=0,
\end{equation}
where  $R(z)$ is an analytic function of  $z$. In order to obtain an approximate solution to Eq. (\ref{eq_ori}), we are going to use the phase--integral   method  developed by Fr\"oman \cite{froman:1965,froman:1966B}. The phase integral approximation, generated using a non specified base solution  $Q(z)$,  is a linear
combination of the phase integral functions \cite{froman:1974A,froman:1996}, which exhibit the following  form

\begin{equation}\label{pi_uk}
u_k=q^{-\sfrac{1}{2}}(z) \exp\left[\pm i\, \omega(z) \right],
\end{equation}
where

\begin{equation}\label{omega}
\omega(z)=\int^z q(z) \, \D z.
\end{equation}
Substituting Eq. (\ref{pi_uk}) into Eq. (\ref{eq_ori}) we obtain that the exact phase integrand $q(z)$ must be a solution of the differential equation,

\begin{equation}
\label{q_ori}	q^{-\sfrac{3}{2}}(z)\dfrac{\D ^2}{\D z^2}q^{-\sfrac{1}{2}}(z)+\dfrac{R(z)}{q^2(z)}-1=0.
\end{equation}
For any solution of Eq. (\ref{q_ori}) the functions (\ref{pi_uk}) are linearly independent, the linear combination of the functions $u_k$ represents a local solution. In order to solve the global problem, we choose a linear combination of phase integral solutions representing the same solution in different regions of the complex plane. This is known as the Stokes phenomenon \cite{froman:1965}.

If we have a function $Q(z)$ which is an approximate solution of Eq. (\ref{q_ori}), the quantity $\epsilon_0$, obtained after substituting $Q(z)$ into Eq. (\ref{q_ori}),

\begin{equation}
\label{epsilon_0}
\epsilon_0=Q^{-\sfrac{3}{2}}(z)\dfrac{\D^2}{\D z^2}Q^{-\sfrac{1}{2}}(z)+\dfrac{R(z)-Q^2(z)}{Q^2(z)},
\end{equation}
is small compared to unity. We take into account the relatively small size of $\epsilon_0$ considering it to be proportional to $\lambda^2$, where $\lambda$ is a small parameter. The parameter $\epsilon_0$ is small when $Q(z)$ is proportional to $1/\lambda$ and $R(z)-Q^2(z)$ is independent of $\lambda$, i.e., if  $R(z)$ is replaced by $Q^2(z)/\lambda^2+\left[R(z)-Q^2(z)\right]$ in Eq. (\ref{eq_ori}). Therefore, instead of considering Eq. (\ref{eq_ori}), we deal with the auxiliary differential equation,

\begin{equation}
\label{eq_aux}
\dfrac{\D^2 u_k}{\D z^2}+\left\{\dfrac{Q^2(z)}{\lambda^2}+\left[R(z)-Q^2(z)\right]\right\}
u_k=0,
\end{equation}
which reduces to Eq. (\ref{eq_ori}) when  $\lambda=1$.
Inserting the solutions (\ref{pi_uk}) into the auxiliary differential equation (\ref{eq_aux}), we obtain the following equation for $q(z)$

\begin{equation}
q^{\sfrac{1}{2}}\dfrac{\D^2}{\D z^2}q^{-\sfrac{1}{2}}-q^2+\dfrac{Q^2(z)}{\lambda^2}+R(z)-Q^2(z)=0,
\end{equation}
which is called the auxiliary $q$ equation.  After introducing the
new variable $\xi$,

\begin{equation}
\xi=\int^zQ(z) \, \D z,
\end{equation}
we obtain,

\begin{equation}
\label{qQdif}
1-\left[\dfrac{q\lambda}{Q(z)}\right]^2+\epsilon_0\lambda^2+\left[\dfrac{q\lambda}{Q(z)}\right]^{\sfrac{1}{2}}\dfrac{\D^2}{\D\xi^2}\left[\dfrac{q\lambda}{Q(z)}\right]^{-\sfrac{1}{2}}\lambda^2=0,
\end{equation}
where $\epsilon_0$ is defined by Eq. (\ref{epsilon_0}). A formal
solution of Eq. (\ref{qQdif}) is obtained after identification,

\begin{equation}\label{qlambdaQ}
\dfrac{q\lambda}{Q}=\sum^\infty_{n=0} Y_{2n}\lambda^{2n}.
\end{equation}

Substituting Eq. (\ref{qlambdaQ}) in Eq. (\ref{qQdif}), we obtain,

\begin{equation}\label{recurrence}
1-\left(\sum_n
Y_{2n}\lambda^{2n}\right)^2+\epsilon_0\lambda^2+\left(\sum_n
Y_{2n}\lambda^{2n}\right)^{\sfrac{1}{2}}\dfrac{\D^2}{\D\xi^2}\left(\sum_n
Y_{2n}\lambda^{2n}\right)^{-\sfrac{1}{2}}=0.
\end{equation}

\vspace{0.25cm}
Using computer manipulation algebra, it is straightforward to obtain
the coefficients $Y_{2n}$. The first values are
\cite{froman:1966B,campbell:1972},

\vspace{-0.25cm}
\begin{eqnarray}
\label{Y0}
Y_0&=&1,\\
\label{Y2}
Y_2&=& \dfrac{1}{2} \,\epsilon_0,\\
Y_4&=&- \dfrac{1}{8} \left(\epsilon_0^2+\epsilon_2 \right),\\
Y_6&=& \dfrac{1}{32} \left(2 \epsilon_0^2 + 6 \epsilon_0\epsilon_2+5\epsilon_1^2+\epsilon_4\right),\\
\end{eqnarray}
where  $\epsilon_\nu$ is defined as,

\begin{equation}\label{epsilon_nu}
\epsilon_\nu=\dfrac{1}{Q(z)}\dfrac{\D \epsilon_{\nu-1}}{\D z}, \quad \nu
\ge 1.
\end{equation}
Truncating the series (\ref{qlambdaQ}) at $n=N$ with $\lambda=1$ we
obtain

\begin{equation}
\label{qQ}
q(z)=Q(z)\sum^N_{n=0} Y_{2n},
\end{equation}
Substituting Eq. (\ref{qQ}) into Eq. (\ref{omega}) we have,

\begin{equation}
\label{omegadef}
\omega(z)=\sum_{n=0}^N \omega_{2n}(z),
\end{equation}
where,

\begin{equation}
\label{omega_sum}
\omega_{2n}(z)=\int^z Y_{2n}Q(z) \, \D z.
\end{equation}

From   Eqs. (\ref{pi_uk}), (\ref{qQ}), and (\ref{omegadef}) we obtain a phase integral approximation of order $2N+1$ generated with the help of the base function $Q(z)$.

The base function $Q(z)$ is not specified and its selection depends on the problem in question. In many cases, it is enough to choose $Q^2(z)=R(z)$, and the first--order phase integral approximation reduces to the WKB approximation. In the first--order approximation, it is convenient to choose a root of $Q^2(z)$ as the lower integration limit in expression (\ref{omega_sum}). However,  for higher orders, i.e., for $2N+1>1$, this is not possible because the function $q(z)$ is singular at the zeros of $Q^2(z)$. In this case, it is convenient to express $\omega_{2n}(z)$ as a contour integral over a two--sheet Riemann surface where $q(z)$ is single valued \cite{froman:1966B}. We define,

\begin{equation}
\omega_{2n}(z)=\dfrac{1}{2} \int_{\Gamma_{t}}Y_{2n}(z)Q(z)\, \D z,
\end{equation}
where  $t$ is a zero of $Q^2(z)$ and  $\Gamma_{t}$ is an integration contour  starting  at the point corresponding to $z$ over a Riemann sheet adjacent to the complex plane, and that encloses the point $t$, in the positive or negative sense and ends at the point $z$.

If the function $Q(z)$ is chosen conveniently, the quantity $\mu$ defined by,

\begin{equation}
\label{validity_phase}
\mu=\mu(z,z_0)=\left|\int_{z_0}^z\left|\epsilon(z)q(z)\,
\D z\right|\right|,
\end{equation}
is much smaller than $1$.  The function $\epsilon(z)$ is given by the left side of the equation (\ref{q_ori}),

\begin{equation}
\epsilon(z)=q^{-\sfrac{3}{2}}(z)\dfrac{\D^2}{\D z^2}q^{-\sfrac{1}{2}}(z)+\dfrac{R(z)}{q^2(z)}-1,
\end{equation}
where the integral $\mu$ measures the accuracy of the phase--integral approximation \cite{froman:2002}.

We assume that the function $Q^2(z)$ is real on the real axis. Taking into account this restriction, we shall call turning point, the zero of $Q^2(z)$. We want to know the connection formulas at both sides of an isolated turning point $z_{ret}$, i.e., a turning point which is located far from other turning points. We will adopt the terms  ``classically permitted region'' and  ``classically forbidden region'' in order to denote those ranges over the real axis where $Q^2(z)>0$ and $Q^2(z)<0$, respectively.
Then, there are two ranges where to define the solution. To the left of the turning point  $0<t<t_\ret$ we have the classically permitted region $Q_{\sca,\ten}^2(k,t)>0$ and to the right of the turning point $t>t_\ret$ corresponding to the classically forbidden region
$Q_{\sca,\ten}^2(k,t)<0$, such as it is shown in Figure \ref{alpha_QSa}  and Figure \ref{alpha_QTa}.

The connection formula for an approximate solution that crosses the turning point $z_{ret}$ from a classically permitted region to a classically forbidden region is \cite{froman:1970A}

\begin{equation}\label{allowed-forbbiden}
\left|
q^{-\sfrac{1}{2}}(z)\right|\cos\left(\left|\omega(z)\right|+\dfrac{\pi}{4}\right)\rightarrow
\left| q^{-\sfrac{1}{2}}(z)\right|\exp\left[\left|\omega(z)\right|\right].
\end{equation}

The connection formula for an approximate solution that crosses the turning point $z_{ret}$ from a classically forbidden region to a classically permitted region is \cite{froman:1970A}

\begin{equation}\label{forbbiden-allowed}
\left| q^{-\sfrac{1}{2}}(z)\right|\exp\left[-\left|\omega(z)\right|\right]
\rightarrow 2 \left|
q^{-\sfrac{1}{2}}(z)\right|\cos\left(\left|\omega(z)\right|-\dfrac{\pi}{4}\right).
\end{equation}

It is important to emphasize the one--directional character of the connection formulas (\ref{allowed-forbbiden}) and (\ref{forbbiden-allowed}), this means that the trace of the solution should be done in the direction indicated by the arrows in the equation (\ref{allowed-forbbiden}) and Eq. (\ref{forbbiden-allowed}).

The equation mode $k$ for the scalar and tensor perturbations (\ref{alpha_ddotUk}) and (\ref{alpha_ddotVk}) in the phase--integral method have two solutions: for $0<t <t_\ret$

\begin{eqnarray}
\label{alpha_uk_left}
u^\phai_k(t)&=& \dfrac{c_1}{\sqrt{a(t)}}\left|q_\sca^{-\sfrac{1}{2}}(k,t)\right| \cos{\left[\left|\omega_\sca(k,t)\right|-\dfrac{\pi}{4}\right]} \\
\nonumber
&+& \dfrac{c_2}{\sqrt{a(t)}}\left|q_\sca^{-\sfrac{1}{2}}(k,t)\right| \cos{\left[\left|\omega_\sca(k,t)\right|+\dfrac{\pi}{4}\right]},\\
\label{alpha_vk_left}
v^\phai_k(t)&=& \dfrac{d_1}{\sqrt{a(t)}}\left|q_\ten^{-\sfrac{1}{2}}(k,t)\right| \cos{\left[\left|\omega_\ten(k,t)\right|-\dfrac{\pi}{4}\right]} \\
\nonumber
&+ &\dfrac{d_2}{\sqrt{a(t)}} \left|q_\ten^{-\sfrac{1}{2}}(k,t)\right| \cos{\left[\left|\omega_\ten(zk,t\right|+\dfrac{\pi}{4}\right]},
\end{eqnarray}

and for $t>t_\ret$

\begin{eqnarray}
\label{alpha_uk_right}
u^\phai_k(t)&=&\dfrac{c_1}{2\sqrt{a(t)}}\left|q_\sca^{-\sfrac{1}{2}}(k,t)\right|\exp\left[-\left|\omega_\sca(k,t)\right|\right]\\
\nonumber
&+& \dfrac{c_2}{\sqrt{a(t)}} \left|q_\sca^{-\sfrac{1}{2}}(k,t)\right| \exp\left[\left|\omega_\sca(k,t)\right|\right],\\
\label{alpha_vk_right}
v^\phai_k(t)&=&\dfrac{d_1}{2\sqrt{a(t)}}\left|q_\ten^{-\sfrac{1}{2}}(k,t)\right|\exp\left[-\left|\omega_\ten(k,t)\right|\right]\\
\nonumber
& +& \dfrac{d_2}{\sqrt{a(t)}} \left|q_\ten^{-\sfrac{1}{2}}(k,t)\right| \exp\left[\left|\omega_\ten(k,t)\right|\right].
\end{eqnarray}
Notice that Eq.  \eqref{limit_uk} and Eq. \eqref{limit_vk} are identical to Eq.  \eqref{alpha_uk_right} and Eq. \eqref{alpha_vk_right}  obtained in the first--order phase--integral   method.

Using the phase--integral  method up to third--order ($2 n+1=3\rightarrow n=1$), we have that $q_{\sca}(k,t)$ and $q_{\ten}(k,t)$ can be expanded in the form,

\begin{eqnarray}
\label{q1}
\nonumber
q_\sca(k,t)&=&\sum_{n=0}^1 Y_{2n_\sca}(k,t) Q_\sca(k,t)\\
         &=&\bigg[Y_{0_\sca}(k,t)+Y_{2_\sca}(k,t)\bigg] Q_\sca(k,t),\\
\label{q2}
\nonumber
q_\ten(k,t)&=&\sum_{n=0}^1 Y_{2n_\ten}(k,t)Q_\ten(k,t)\\
&=&\bigg[Y_{0_\ten}(k,t)+Y_{2_\ten}(k,t) \bigg] Q_\ten(k,t).
\end{eqnarray}

In order to compute $q_{\sca,\ten}(k,t)$, we calculate $Y_{2_{\sca,\ten}}(k,t)$ and the required function $\varepsilon_{0_{\sca,\ten}}(k,t)$.  The expressions (\ref{q1}) and \eqref{q2} give a third--order approximation for $q_{\sca,\ten}(k,t)$.  In order to compute $\omega_{\sca,\ten}(k,t)$  we make a contour integration following the path indicated in Figures \ref{alpha_QTb}--(c),

\begin{eqnarray}
\nonumber
\omega_\sca(k,t)&=&\omega_{0_\sca}(k,t)+ \omega_{2_\sca}(k,t)\\
\nonumber
&=&\int_{\tau_\sca}^{t}Q_\sca(k,t)\,\D t+\dfrac{1}{2}\int_{\Gamma_{\tau_\sca}}Y_{2_\sca}(k,t)Q_\sca(k,t)\,\D t\\
&=&\int_{\tau_\sca}^{t}Q_\sca(k,t)\,\D t+\dfrac{1}{2}\int_{\Gamma_{\tau_\sca}}f_{2_\sca}(k,t)\,\D t.
\end{eqnarray}

\begin{eqnarray}
\nonumber
\omega_\ten(k,t)&=&\omega_{0_\ten}(k,t)+ \omega_{2_\ten}(k,t)\\
\nonumber
&=&\int_{\tau_\ten}^{t}Q_\ten(k,t)\,\D t+\dfrac{1}{2}\int_{\Gamma_{\tau_\ten}}Y_{2_\ten}(k,t)Q_\ten(k,t)\, \D t\\
&=&\int_{\tau_\sca}^{t}Q_\ten(k,t)\,\D t+\dfrac{1}{2}\int_{\Gamma_{\tau_\ten}}f_{2_\ten}(k,t)\,\D t,
\end{eqnarray}

where
\begin{eqnarray}
f_{2_\sca}(k,t)&=&Y_{2_\sca}(k,t)Q_\sca(k,t),\\
f_{2_\ten}(k,t)&=&Y_{2_\ten}(k,t)Q_\ten(k,t).
\end{eqnarray}
The functions  $f_{2_\sca}(k,t)$ and  $f_{2n_\ten}(k,t)$ have the following functional dependence:

\begin{eqnarray}
\label{A}
f_{2_\sca}(k,t)&=&A(k,t)(t-\tau_\sca)^{\sfrac{-5}{2}},\\
\label{B}
f_{2_\ten}(k,t)&=&B(k,t)(t-\tau_\ten)^{\sfrac{-5}{2}},
\end{eqnarray}
where the functions $A(k,t)$ are regular in $\tau_\sca$, and the function $B(k,t)$ are regular in $\tau_\ten$. With the help of the functions \eqref{A}-\eqref{B} we compute the integrals for $\omega_{2_{\sca,\ten}}$ using the contour indicated in Figs. \ref{alpha_QSb}--(c) and \ref{alpha_QTb}--(c). The expressions for $\omega_{2_{\sca,\ten}}$ permit one to obtain the third-order phase integral approximation of the solution to the equations for scalar \eqref{alpha_ddotUk} and tensor \eqref{alpha_ddotVk} perturbations.  The constants $c_1$, $c_2$, $d_1$ and $d_2$ are obtained using the limit $k\,t\rightarrow 0$ of the solutions on the left side of the turning point \eqref{alpha_uk_left} and \eqref{alpha_vk_left}, and are given by the expressions.

\begin{eqnarray}
c_1&=&-\im\,c_2,\\
c_2&=&\dfrac{\e^{-\im\frac{\pi}{4}}}{\sqrt{2}}\e^{-\im\left[k\,\eta(0)+\left|\omega_{0_\sca}(k,0)\right|\right]},\\
d_1&=&-\im\,d_2,\\
d_2&=&\dfrac{\e^{-\im\frac{\pi}{4}}}{\sqrt{2}}\e^{-\im\left[k\,\eta(0)+\left|\omega_{0_\ten}(k,0)\right|\right]}.
\end{eqnarray}
In order to compute the scalar and tensor power spectra, we need to calculate the limit as $k\,t\rightarrow \infty$ of the growing part of the solutions on the right side of the turning point given by Eq. \eqref{alpha_uk_right} and  Eq. \eqref{alpha_vk_right} for scalar and tensor perturbations, respectively.

\begin{eqnarray}
\label{PS_pi}
P^\phai_\sca(k)&=&\lim_{-k t\rightarrow \infty} \dfrac{k^3}{2\pi^2} \left|\dfrac{u_k^\phai(t)}{z_\sca(t)}\right|^2,\\
\label{PT_pi}
P^\phai_\ten(k)&=&\lim_{-k t\rightarrow \infty} \dfrac{k^3}{2\pi^2} \left|\dfrac{v_k^\phai(t)}{a(t)}\right|^2,
\end{eqnarray}
where `$\phai$' means the third--order phase integral method.

We also compared the performance of the methods with both models: $\alpha$--attractor and Starobinsky inflationary model. We calculated, using numerical integration and approximations detailed previously, observable quantities, and obtained a quantitative measurement of their performance in each case.  We found that the calculation behaves better for the analysis of the scalar spectral index $n_{\sca}(k)$ of Starobinsky potential, this is a lower relative error in all the approximation methods used. Also, $\alpha$--attractor offers a lower value for the tensor scalar ratio $r(k)$. The observable parameters obtained here for $\alpha$--attractor inflation
and Starobinsky inflation are within the range of Planck's data as expected.


\bibliographystyle{unsrt}

\begin{thebibliography}{10}

\bibitem{linde1984inflationaryONE}
{A. D. Linde}.
\newblock {The inflationary universe}.
\newblock {\em Rep. Prog. Phys.}, 47:925, 1984.

\bibitem{baumann2009tasiTWO}
{D. Baumann}.
\newblock {TASI lectures on inflation}.
\newblock {\em arXiv preprint arXiv:0907.5424}, 2009.

\bibitem{martin2017theoryFOUR}
{J. Martin}.
\newblock {The Theory of Inflation, in proceedings of the 200th Course of
  Enrico Fermi School of Physics: Gravitational Waves and Cosmology
  (GW--COSM)}, 2017.

\bibitem{martin:2014}
{J. Martin, C. Ringeval, and V. Vennin}.
\newblock {Encyclopaedia Inflationaris}.
\newblock {\em Phys. Dark Univ.}, 5--6:75, 2014.

\bibitem{riess1998observational}
{A. G. Riess, \textit{et al.}}
\newblock {Observational evidence from supernovae for an accelerating universe
  and a cosmological constant}.
\newblock {\em AJ}, 116:1009, 1998.

\bibitem{Perlmutter1999}
{S. Perlmutter, \textit{et al.}}
\newblock {Measurements of $\Omega$ and $\Lambda$ from 42 High--Redshift
  Supernovae}.
\newblock {\em ApJ}, 517:565, 1999.

\bibitem{Li2013}
{M. Li, X. Li, S. Wang, and Y. Wang}.
\newblock Dark energy.
\newblock {\em Front. Phys.}, 8:828, 2013.

\bibitem{Abbott2023}
T.~M.~C. Abbott and et~al.
\newblock Dark energy survey year 3 results: Constraints on extensions to
  {$\Lambda$CDM} with weak lensing and galaxy clustering.
\newblock {\em Phys. Rev. D}, 107:083504, 2023.

\bibitem{Peebles1988}
{P. J. E. Peebles, and B. Ratra}.
\newblock Cosmology with a time variable cosmological ``constant''.
\newblock {\em ApJ}, 325:L17, 1988.

\bibitem{Coble1997}
{K. Coble, S. Dodelson, and J. Frieman}.
\newblock Dynamical $\lambda$ models of structure formation.
\newblock {\em Phys. Rev. D}, 55:1851, 1997.

\bibitem{Lopez1996}
J.~L. López and D.~V. Nanopoulos.
\newblock A new cosmological constant model.
\newblock {\em Mod. Phys. Lett. A}, 11:1, 1996.

\bibitem{Linde1995}
{A. D. Linde}.
\newblock Inflation with variable $\omega$.
\newblock {\em Phys. Lett. B}, 351:99, 1995.

\bibitem{Wetterich1988}
{C. Wetterich}.
\newblock Cosmology and the fate of dilatation symmetry.
\newblock {\em Nucl. Phys. B}, 302:668, 1988.

\bibitem{Caldwell1998}
{R. R. Caldwell, R. Dave, P. J. Steinhardt}.
\newblock Cosmological imprint of an energy component with general equation of
  state.
\newblock {\em Phys. Rev. Lett.}, 80:1582, 1998.

\bibitem{Tsujikawa2013}
{S. Tsujikawa}.
\newblock Quintessence: A review.
\newblock {\em Class. Quantum Grav.}, 30:214003, 2013.

\bibitem{PeeblesVilenkin1999}
{P. J. E Peebles, and A. Vilenkin}.
\newblock Quintessential inflation.
\newblock {\em Phys. Rev. D}, 59:063505, 1999.

\bibitem{Sahni2001}
{V. Sahni, M. Sami, and T. Souradeep}.
\newblock Relic gravity waves from brane world inflation.
\newblock {\em Phys. Rev. D}, 65:023518, 2001.

\bibitem{DimopoulosValle2002}
{K. Dimopoulos, and J. W. F. Valle}.
\newblock Modeling quintessential inflation.
\newblock {\em Astropart. Phys.}, 18:287, 2002.

\bibitem{Hossain2015}
{Md. Wali Hossain, R. Myrzakulov, M. Sami, and E. N. Saridakis}.
\newblock Unification of inflation and dark energy \textit{à la}
  quintessential inflation.
\newblock {\em Int. J. Mod. Phys. D}, 24:1530014, 2015.

\bibitem{deHaro2021}
{L. A. Saló, D. Benisty, E. I. Guendelman, and J. d. Haro}.
\newblock Quintessential Inflation and Cosmological Seesaw Mechanism: Reheating and Observational Constraints.
\newblock {\em JCAP}, 07:007, 2021.

\bibitem{BettoniRubio2022}
{D. Bettoni, and J. Rubio}.
\newblock Quintessential inflation: A tale of emergent and broken symmetries.
\newblock {\em Galaxies}, 10:22, 2022.

\bibitem{Kallosh_Linde:2013b}
{R. Kallosh, and A. Linde}.
\newblock {Universality class in conformal inflation}.
\newblock {\em JCAP}, 2013:002, 2013.

\bibitem{Kallosh_Linde_Roest:2013}
{R. Kallosh, A. Linde, and D. Roest}.
\newblock {Superconformal Inflationary $\alpha$--attractors}.
\newblock {\em JHEP}, 2013:198, 2013.

\bibitem{galante}
{M. Galante, R. Kallosh, A. Linde, and D. Roest}.
\newblock Unity of cosmological inflation attractors.
\newblock {\em Phys. Rev. Lett}, 114:141302, 2015.

\bibitem{carrasco2015}
{J. J. M. Carrasco, R. Kallosh, and A. Linde}.
\newblock {Cosmological attractors and initial conditions for inflation}.
\newblock {\em Phys. Rev. D}, 92:063519, 2015.

\bibitem{carrasco2016}
{J. J. M. Carrasco, R. Kallosh, and A. Linde}.
\newblock {Minimal supergravity inflation}.
\newblock {\em Phys. Rev. D}, 93, 2016.

\bibitem{ferrara}
{S. Ferrara, R. Kallosh, A. Linde, and M. Porrati}.
\newblock {Minimal supergravity models of inflation}.
\newblock {\em Phys. Rev. D}, 88:085038, 2013.

\bibitem{kallosh8}
Renata Kallosh, Andrei Linde, Diederik Roest, Alexander Westphal, and Yusuke
  Yamada.
\newblock Fibre inflation and $\alpha$--attractors.
\newblock {\em JHEP}, 2018:1, 2018.

\bibitem{akrami:2020}
Y.~Akrami \textit{et al.}
\newblock {Planck 2018 results. X. Constraints on inflation}.
\newblock {\em Astron. Astrophys.}, 641:A10, 2020.

\bibitem{rodrigues2021observational}
{J. G. Rodr\'iguez, S. Santos da Costa, and J. S. Alcaniz}.
\newblock {Observational constraints on $\alpha$--attractor inflationary models
  with a Higgs--like potential}.
\newblock {\em Phys. Lett. B}, 815:136156, 2021.

\bibitem{bhatta}
{S. Bhattacharya, K. Dutta, M. R. Gangopadhyay, and A. Maharana}.
\newblock {$\alpha$--attractor inflation: Models and predictions}.
\newblock {\em Phys. Rev. D}, 107:103530, 2023.

\bibitem{PettorinoAmendolaWetterich2013}
{V. Pettorino, L. Amendola, and C. Wetterich}.
\newblock {How early is early dark energy?}
\newblock {\em Phys. Rev. D}, 87:083009, 2013.

\bibitem{KarwalKamionkowski2016}
Tanvi Karwal and Marc Kamionkowski.
\newblock Dark energy at early times, the hubble parameter, and the string
  axiverse.
\newblock {\em Phys. Rev. D}, 94:103523, 2016.

\bibitem{PoulinSmithKarwalKamionkowski2019}
{V. Poulin, T. L. Smith, T. Karwal, and M. Kamionkowski}.
\newblock Early dark energy can resolve the hubble tension.
\newblock {\em Phys. Rev. Lett.}, 122:221301, 2019.

\bibitem{sabla}
{V. I. Sabla, and R. R. Caldwell}.
\newblock Microphysics of early dark energy.
\newblock {\em Phys. Rev. D}, 106:063526, 2022.

\bibitem{BragliaEmondFinelliGumrukcuogluKoyama2020}
{M. Braglia, W. T. Emond, F. Finelli, A. Emir Gumrukcuoglu, and K. Koyama}.
\newblock {Unified framework for early dark energy from $\alpha$--attractors}.
\newblock {\em Phys. Rev. D}, 102:083513, 2020.

\bibitem{BrissendenDimopoulosSanchezLopez2024}
{L. Brissenden, K. Dimopoulos, and S. S\'anchez L\'opez}.
\newblock {Non--oscillating early dark energy and quintessence from
  $\alpha$--attractors}.
\newblock {\em Astropart. Phys.}, 157:102925, 2024.

\bibitem{sarkar}
{A. Sarkar, and B. Ghosh}.
\newblock {Constraining an Early Dark Energy Motivated Quintessential
  $\alpha$--Attractor Inflaton Potential}.
\newblock {\em arXiv preprint arXiv:2307.00603}, 2024.

\bibitem{salo:2021}
{L. A. Sal\'o, D. Benisty, E. I. Guendelman, and. J. d. Haro}.
\newblock {$\alpha$--attractors in quintessential inflation motivated by
  supergravity}.
\newblock {\em Phys. Rev. D}, 103:123535, 2021.

\bibitem{akrami2021quintessential}
{Y. Akrami, S. Casas, S. Deng, and V. Vardanyan}.
\newblock {Quintessential $\alpha$--attractor inflation: forecasts for Stage IV
  galaxy surveys}.
\newblock {\em JCAP}, 2021:006, 2021.

\bibitem{linder2015dark}
{E. V. Linder}.
\newblock {Dark energy from $\alpha$--attractors}.
\newblock {\em Phys. Rev. D}, 91:123012, 2015.

\bibitem{giare2024testing}
{W. Giar{\`e}, E. Di Valentino, E. V. Linder and E. Specogna}.
\newblock {Testing $\alpha$--attractor quintessential inflation against CMB and
  low--redshift data}.
\newblock {\em Phys. Dark. Univ.}, 46:101713, 2024.

\bibitem{zhumabek2023connecting}
{T. Zhumabek, M. Denissenya, and E. V. Linder}.
\newblock {Connecting primordial gravitational waves and dark energy}.
\newblock {\em JCAP}, 2023, 2023.

\bibitem{de2008calibrating}
{R. De Putter, and E. V. Linder}.
\newblock {Calibrating dark energy}.
\newblock {\em JCAP}, 2008:042, 2008.

\bibitem{yadav2011dark}
{A. K. Yadav, F. Rahaman, and S. Ray}.
\newblock {Dark energy models with variable equation of state parameter}.
\newblock {\em IJTP}, 50:871, 2011.

\bibitem{liddle2000cosmological}
{A. R. Liddle, and D. H. Lyth}.
\newblock {\em {Cosmological inflation and large--scale structure}}.
\newblock Cambridge university press, 2000.

\bibitem{kallosh3}
{R. Kallosh, and A. Linde}.
\newblock {Cosmological attractors and asymptotic freedom of the inflaton
  field}.
\newblock {\em JCAP}, 2016:047, 2016.

\bibitem{Kallosh_Linde:2013}
{R. Kallosh, and A. Linde}.
\newblock {Superconformal generalizations of the Starobinsky model}.
\newblock {\em JCAP}, 2013:028, 2013.

\bibitem{canas:2021}
{G. Ca\~nas--Herrera, and F. Renzi}.
\newblock {Current and Future constraints on single--field $\alpha$--attractor
  model}.
\newblock {\em Phys. Rev. D}, 104:103512, 2021.

\bibitem{Miranda_Fabris_Piattella:2017}
{T. Miranda, J. C. Fabris, and O. F. Piattella}.
\newblock {Reconstructing a $f(R)$ theory from the $\alpha$--attractors}.
\newblock {\em JCAP}, 2017:041, 2017.

\bibitem{truman:2020}
{T. Tapia, M. Z. Mughal, and C. Rojas}.
\newblock {Semiclassical analysis of the Starobinsky inflationary model}.
\newblock {\em Phys. Dark Univ.}, 30:100650, 2020.

\bibitem{hinshaw2013nineDOCE}
{G. Hinshaw \textit{et al.}}
\newblock {Nine--year Wilkinson Microwave Anisotropy Probe (WMAP) observations:
  cosmological parameter results}.
\newblock {\em ApJS}, 208:19, 2013.

\bibitem{ade2016planckTRECE}
{P. A. R. Ade, \textit{et al.}}
\newblock {Planck 2015 results--XXVIII. the planck catalogue of galactic cold
  clumps}.
\newblock {\em A\&A}, 594:A28, 2016.

\bibitem{iacconi2023novelQUINCE}
{L. Laconi, M. Fasiello, J. V\"aliviita, and D. Wands}.
\newblock {Novel CMB constraints on the $\alpha$ parameter in alpha--attractor
  models}.
\newblock {\em JCAP}, 2023:015, 2023.

\bibitem{iacconi2025testingDISIX}
{L. Iacconi, M. Bacchi, L. P. Guimar\~aes, and F. T. Falciano}.
\newblock {Testing inflation on all scales: a case study with
  $\alpha$--attractors}.
\newblock {\em JCAP}, 2025:004, 2025.

\bibitem{alestas2025desiSEVEN}
{G. Alestas, M. Caldarola, S. Kuroyanagi, and S. Nesseris}.
\newblock {DESI constraints on $\alpha$--attractor inflationary models}.
\newblock {\em Phys. Rev. D}, 111:083506, 2025.

\bibitem{Mathematica:2025}
{Wolfram Research, Inc.}
\newblock Mathematica, version 14.3.0.0.
\newblock \url{https://www.wolfram.com/mathematica/}, 2025.

\bibitem{gnuplot:2024}
{T. Williams \textit{et al.}}
\newblock Gnuplot, version 6.0.
\newblock \url{http://www.gnuplot.info}, 2024.

\bibitem{habib:2005b}
{S. Habib, A. Heinen, K. Heitmann and G. Jungman}.
\newblock {Inflationary Perturbations and Precision Cosmology}.
\newblock {\em Phys. Rev. D}, 71:043518, 2005.

\bibitem{giare:2023b}
{W. Giar\`e, S. Pan, E. Di Valentino, W. Yang, J. de Haro, and A. Melchiorri}.
\newblock {Inflationary Potential as seen from Different Angles: Model
  Compatibility from Multiple CMB Missions}.
\newblock {\em JCAP}, 09:019, 2023.

\bibitem{vazquez:2013}
{J. A. Vazquez, M. Bridges, Y--Z. Ma, and M. P. Hobson}.
\newblock {Constraints on the tensor--to--scalar ratio for non--power--law
  models}.
\newblock {\em JCAP}, 08:001, 2013.

\bibitem{das:2023}
{S. Das and R. O. Ramos}.
\newblock {Running and Running of the Running of the Scalar Spectral Index in
  Warm Inflation}.
\newblock {\em Universe}, 9:76, 2023.

\bibitem{vazquez:2020}
{J. A. Vazquez, L. E. Padilla, and T. Matos}.
\newblock {Inflationary cosmology: from theory to observations}.
\newblock {\em Rev. Mex. Fis. E}, 17:73, 2020.

\bibitem{finelli:2018}
{F. Finelli {\textit et al.}}
\newblock {Exploring cosmic origins with CORE: Inflation}.
\newblock {\em JCAP}, 04:016, 2018.

\bibitem{adshead:2011}
{P. Adshead, R. Easther, J. Pritchard, and A. Loeb}.
\newblock { Inflation and the scale dependent spectral index: prospects and
  strategies}.
\newblock {\em JCAP}, 02:021, 2011.

\bibitem{stewart:1993}
{E. D Stewart and D. H. Lyth}.
\newblock {A more accurate analytic calculation of the spectrum of cosmological
  perturbations produced during inflation}.
\newblock {\em Phys. Lett. B}, 302:171, 2001.

\bibitem{berry:1972}
{M. Berry and K. E. Mount}.
\newblock {Semiclassical Approximations in Wave Mechanics}.
\newblock {\em Rep. Prog. Phys.}, 35:315, 1972.

\bibitem{habib:2002}
{S. Habib, A. Heinen, K. Heitmann, G. Jungman and C. Molina--Par\'is}.
\newblock {The Inflationary Perturbation Spectrum}.
\newblock {\em Phys. Rev. Lett.}, 89:281301, 2002.

\bibitem{rojas:2007b}
{C. Rojas and V. M. Villalba}.
\newblock {Computation of inflationary cosmological perturbations in the
  power--law inflatioary model using the phase--integral method}.
\newblock {\em Phys. Rev. D}, 75:063518, 2007.

\bibitem{rojas:2007c}
{V. M. Villalba and C. Rojas}.
\newblock {Applications of the phase integral method ins ome inflationary
  scenarios}.
\newblock {\em J. Phys. Conf. Ser.}, 66:012034, 2007.

\bibitem{dingle}
{R. B. Dingle}.
\newblock {The method of comparison equations in the solution of linear
  second--order differential equations (generalized WKB method)}.
\newblock {\em Appl. Sci. Res. B}, 5:345, 1956.

\bibitem{abramowitz:1965}
M.~Abramowitz and I.~A. Stegun.
\newblock {\em Handbook of Mathematical Functions}.
\newblock Dover, New York, 1965.

\bibitem{froman:1965}
N.~Fr\"oman and P.~O. F\"oman.
\newblock {\em JWKB Approximation. Contribution to the Theory}.
\newblock North--Holland, Amsterdam, 1965.

\bibitem{froman:1966B}
N.~Fr\"oman.
\newblock {Detailed analysis of some properties of the JWKB--approximation}.
\newblock {\em Ark. Fys.}, 31:381, 1966.

\bibitem{froman:1974A}
N.~Fr\"oman and P.~O. F\"oman.
\newblock A direct method for modifying certain phase--integral approximations
  of arbitrary order.
\newblock {\em Ann. Phys.}, 83:103, 1974.

\bibitem{froman:1996}
N.~Fr\"oman and P.~O. F\"oman.
\newblock {\em {Phase--Integral Method. Allowing Nearlying Transition Point}},
  volume~40.
\newblock Springer Tracts in Natural Philosophy, 1996.

\bibitem{campbell:1972}
J.~A. Campbell.
\newblock Computation of a class of functions useful in the phase--integral
  approximation. \textnormal{I}. \textnormal{Results}.
\newblock {\em J. Comp. Phys.}, 10:308, 1972.

\bibitem{froman:2002}
N.~Fr\"oman and P.~O. F\"oman.
\newblock {\em Physical Problems Solved by the Phase--Integral Method}.
\newblock Cambridge University Press, 2002.

\bibitem{froman:1970A}
N.~Fr\"oman.
\newblock Connection formulas for certain higher order phase--integral
  approximations.
\newblock {\em Ann. Phys.}, 61:451, 1970.

\end{thebibliography}

\end{document}